\documentclass[journal, onecolumn, 12pt]{IEEEtran}
\usepackage{lineno}
\usepackage{setspace}
\usepackage{hyperref}
\usepackage{amsmath}
\usepackage{amssymb}
\usepackage{array}
\usepackage[dvipsnames]{xcolor}
\usepackage[pdftex]{graphicx}
\usepackage{tabularray}
\usepackage{booktabs}
\usepackage{rotating}
\usepackage{arydshln}
\usepackage{pifont}
\usepackage{multirow}
\usepackage{caption} 
\usepackage{subcaption}
\usepackage{float}
\usepackage{algorithm}
\usepackage{algpseudocode}
\usepackage{xcolor}
\usepackage{caption}
\DeclareCaptionFormat{ieeetable}{
TABLE~\thetable\\
\MakeUppercase{#2#3}
}
\usepackage[
  backend=biber,
  style=ieee,
  doi=true,
  url=true,
  isbn=false
]{biblatex}
\renewbibmacro*{urldate}{}
\AtEveryBibitem{%
  \clearfield{urldate}%
}
\begin{document}
\title{Image Compression Using Novel View Synthesis Priors}

\author{Luyuan~Peng,~\IEEEmembership{Member,~IEEE,}
        Mandar~Chitre,~\IEEEmembership{Fellow,~IEEE,}
        Hari~Vishnu,~\IEEEmembership{Senior Member,~IEEE,}
        Yuen~Min~Too,~\IEEEmembership{Member,~IEEE,}
        Bharath~Kalyan,~\IEEEmembership{Senior Member,~IEEE,}
        Rajat~Mishra,~\IEEEmembership{Member,~IEEE}
        and~Soo~Pieng~Tan% <-this % stops a space
        }

\maketitle

% \doublespacing
\begin{abstract}
Real-time visual feedback is essential for tetherless control of remotely operated vehicles, particularly during inspection and manipulation tasks. Though acoustic communication is the preferred choice for medium-range communication underwater, its limited bandwidth renders it impractical to transmit images or videos in real-time. To address this, we propose a model-based image compression technique that leverages prior mission information. Our approach employs trained machine-learning based novel view synthesis models to store and render views of scenes that serve as priors, and minimizes the differences between camera images and model-rendered images to improve compressibility. We evaluate our method using datasets collected in an artificial ocean basin and demonstrate significant improvements in compression ratio and image quality compared to existing methods. Moreover, our approach exhibits robustness to the introduction of novel objects within the basin scene and maintains performance under real-world degradations such as backscatter, as demonstrated on an online coral reef dataset. These results highlight the promise of our proposed technique for enabling real-time, high-quality visual feedback over bandwidth-constrained underwater acoustic links.
\end{abstract}

% \begin{IEEEkeywords}
% underwater, image compression, 3D gaussian splatting, remotely operated vehicle, acoustic communication.
% \end{IEEEkeywords}

\IEEEpeerreviewmaketitle

\section{Introduction}
\IEEEPARstart{U}{nderwater} missions, such as deep-sea exploration, environmental monitoring, and infrastructure inspection and manipulation, heavily rely on remotely operated vehicles (ROVs)~\cite{vargas_robust_2021,bingham_robotic_2010,carreras_online_2016}. Traditional ROVs are tethered to a surface platform, which provides a continuous power supply, enabling extended operation. The tether also serves as a reliable two-way communication conduit, transmitting control commands from the operator topside unit to the ROV and sending sensor data, video feed, and other information back to the operators~\cite{shepherd_remotely_2001}. While essential, the tether introduces significant limitations. It restricts the ROV's maneuverability, limiting its ability to navigate through tight spaces or complex underwater environments~\cite{bowen_untethered_2013}. This constraint can hinder the ROV from reaching certain areas or performing delicate tasks. Additionally, the tether poses a significant risk of entanglement with underwater structures and obstacles such as rocks or reefs, potentially trapping the ROV or complicating retrieval, which could lead to mission failure or loss of the vehicle. Moreover, the tether’s weight and bulk add to the logistical challenges of deploying and recovering the ROV. This often necessitates additional equipment, such as winches and cable reels, further increasing the operational complexity~\cite{kalyan_concept_2023}.

To mitigate these challenges, the development of tetherless ROVs has been a focus in marine robotics~\cite{bowen_untethered_2013}, with one of the greatest challenges being real-time wireless underwater communication between the ROV and the surface platform. Radio waves provide high data rates in terrestrial environments but suffer severe attenuation in water; as a result, practical underwater links rely on acoustics, with optics used opportunistically at short range~\cite{tan_survey_2011, stojanovic_underwater_2009}.

Underwater wireless links are realized primarily via two modalities: acoustic and optical. Acoustic links provide long range at low power but with limited bandwidth~\cite{stojanovic_underwater_2009}. Optical links can achieve up to about 10 megabits per second yet are short-range, highly directional, and degrade in turbid water~\cite{underwater_optical}. Prior work has therefore followed two directions: increasing acoustic throughput~\cite{mandar_underwater_acoustic} and employing optical links where geometry and water clarity permit~\cite{underwater_optical}. In practice, acoustics remain the most widely usable option, delivering on the order of tens of kilobits per second (kbps) over hundreds of meters to a few kilometers, depending on channel conditions~\cite{acoustics_sota, acoustics_sota2}. Such capacity suffices for commands and low-rate telemetry but is inadequate for real-time image transmission needed for effective teleoperation and inspection. 

Classical image codecs such as WebP~\cite{compression:webp} and JPEG-XL~\cite{compression:jpegxl} are not designed with such constraints in mind and do not offer sufficient compression for such applications. For example, a 320$\times$180 RGB image compressed using WebP is still about 7 kilobytes (kB), allowing a 100kbps link to transmit at most 2 frames per second—often too low for effective teleoperation. Recent learning-based image compression methods have demonstrated superior performance over classical codecs by learning compact, content-adaptive latent representations~\cite{balle_variational, yasin_image_2021, mbt18, mlic, mlic}. However, these approaches typically require large and diverse training datasets to generalize well—resources that are often unavailable in underwater domains.

Underwater inspection missions are frequently conducted at the same sites over time to monitor structural degradation, biofouling, or other environmental changes~\cite{nauert_inspection_2023, ammon_impact_2018}. These repeat visits provide valuable scene-specific prior information, which can be leveraged to optimize image compression. An efficient way to store this information for future reference is to train a Novel View Synthesis (NVS) model. Once trained, the NVS model can be queried at runtime to generate photorealistic views of the scene from arbitrary camera poses, serving as a scene-specific prior for compression. 

With such a prior available, most of the image content can be reconstructed easily, and only the residual differences required to describe the real scene accurately needs to be transmitted by the ROV to the operator. To understand the benefit of this approach, we consider an information-theoretic view: without prior information, the image \( I \) visible to the ROV at an underwater site can be modeled as a sample from a distribution \( p(I) \). The compressed size achievable with this image would be bounded by its entropy \( H(I) = -\mathbb{E}[\log \left(p(I)\right)] \), where $\mathbb{E}$ denotes the expectation operator. When prior knowledge of the scene is available, such as in the form of a prior view \( I_\text{NVS} \) synthesized using an NVS model, this allows us to model the image conditionally as \( p(I \mid I_\text{NVS}) \), which yields a lower entropy \( H(I \mid I_\text{NVS}) \leq H(I) \). In an extreme ideal case, where the scene remains unchanged and \(I = I_\text{NVS}\), the residual is zero, and hence no additional information needs to be transmitted apart from the minimal information required to enable the NVS to render the scene itself. In general, the more informative the prior, the more predictable the scene becomes, and the tighter the lower bound for compression.

In this paper, we propose a novel image compression framework that leverages prior scene knowledge encoded in an NVS model and achieves efficient compression by refining compact latent representations at runtime. Unlike conventional codecs or learned methods trained on large-scale data, our approach adapts to specific environments and achieves high compression ratios while preserving image fidelity and ensuring computational efficiency in bandwidth-limited underwater scenarios. The key contributions of this work are:

\begin{enumerate}  
    \item We introduce \textbf{NVSPrior}, the first image compression framework that exploits scene-specific priors from trained NVS models. This is the first application of NVS priors to image compression, and we systematically validate its effectiveness across both controlled and real-world underwater environments.  
    \item We propose \textbf{iNVS}, a gradient-based latent refinement method that significantly enhances the compression efficiency of NVSPrior while maintaining high reconstruction fidelity and low per-frame latency.  
    \item We conduct a comprehensive analysis of loss functions, optimization algorithms, and initialization strategies, offering practical insights into configuring robust iNVS-based compression for different deployment conditions.  
    \item We demonstrate robustness in (i) a controlled ocean-basin setup, (ii) scenarios with novel objects and occlusions, and (iii) real-world underwater datasets characterized by turbidity, marine snow and backscatter. In all cases, our method outperforms classical and learned baselines in both compression ratio and reconstruction quality.  
\end{enumerate}  

In this work, we focus specifically on source coding for underwater image transmission. Channel impairments such as packet loss, bit errors, and latency are assumed to be handled by existing acoustic modem and protocol stacks, and are therefore outside the scope of this paper.

The paper is organized as follows. Section~\ref{sec:related-works} reviews existing image compression techniques and introduces NVS models, along with prior work on model inversion for pose refinement. Section~\ref{sec:methods} provides an overview of our proposed compression framework and details the optimization strategy. Section~\ref{sec:experiments} describes the datasets, presents ablation studies for model configuration, and demonstrates the robustness and performance of our approach in both controlled and real-world underwater environments. Section~\ref{sec:discussion} discusses the limitations of our method, practical considerations for real-world deployment, and directions for future research. Finally, Section~\ref{sec:conclusion} summarizes the main findings.

\section{Related Work}
\label{sec:related-works}
\subsection{Traditional Image Codecs}
Image compression has long been essential for efficient storage and transmission. 
Traditional image compression methods rely on hand-crafted transformations and statistical encoding techniques to efficiently represent image data. JPEG-XL, the most recent and advanced format in the JPEG family, combines discrete cosine transform-based block coding (similar to traditional JPEG) with Haar wavelet transforms, enabling high compression efficiency and scalability~\cite{compression:jpegxl}. WebP, a widely adopted format optimized for web applications, employs block-based predictive coding for lossy compression and Huffman coding for entropy encoding, making it well-suited for fast decoding and reduced file sizes~\cite{compression:webp}. These formats balance efficiency, scalability, and practical application across various digital environments. While effective for general-purpose use, these formats struggle to meet the constraints of low-bandwidth underwater acoustic links, especially for high-resolution or real-time transmission.
    
\subsection{Learned Image Compression}
End-to-end learned image compression uses neural networks to learn content-adaptive latent representations and now matches or surpasses classical codecs such as JPEG 2000 and WebP in rate–distortion (RD) performance~\cite{balle_variational, toderici_full_2017, whang_neural_2022, gregor_towards_2016, sento_image_2016, theis_lossy_2017, mbt18, cheng2020, mlic}. The field has progressed from hyperprior models that capture spatial dependencies in the latents~\cite{balle_variational, mbt18}, to richer context models that further improve RD at the cost of higher complexity~\cite{cheng2020, mlic}, and most recently to MLIC++, which attains state-of-the-art RD performance and efficiency via linear-complexity multi-reference entropy modeling~\cite{mlic}. However, generic learned image compression models typically require large, diverse training data—scarce in underwater inspection—and lack mechanisms to exploit scene-specific priors available in structured, repeat-survey environments. For benchmarking in this work, we adopt the Mean \& Scale Hyperprior method \cite{mbt18} for its real-time performance, and use MLIC++ to represent the state of the art~\cite{mlic}.

\subsection{NVS-based Compression}
NVS is the process of generating images of a scene from viewpoints that were not originally captured. 3D Gaussian splatting (3DGS)~\cite{kerbl_3dgs_2023}, a recent advancement in NVS techniques, presents a new possibility for real-time image compression. 3DGS models a scene using explicit 3D point-based primitives with Gaussian attributes, and performs rendering using a differentiable splatting process, where Gaussians are projected onto the image plane and blended using alpha compositing. A 3DGS model can be trained using a sparse set of 2D images and offers real-time, highly photorealistic rendering~\cite{kerbl_3dgs_2023}.

Several recent efforts have also examined the use of NVS models for compression tasks. Zhang et al. introduce 2D Gaussian splatting for speeding up encoding and decoding of individual images~\cite{2d_gaussian}, while Liu et al. (2024) propose HEMGS, a hybrid entropy model designed for compressing 3DGS data which are the Gaussian attributes~\cite{hemgs}. In contrast, our method uses gradient descent through a trained NVS model to achieve compression of camera images for real-time transmission over limited-bandwidth acoustic links.

\subsection{Existing Work in Inverting NVS Models}
Recent work has explored gradient descent optimization through trained NVS models, such as Neural Radiance Fields (NeRF) and 3DGS, for camera pose estimation and refinement~\cite{yen-chen_inerf_2021,sun_icomma_2024, gscpr,gsloc,featloc}. For instance, iNeRF~\cite{yen-chen_inerf_2021} estimates the camera pose by inverting a single image using a pre-trained NeRF model, minimizing pixel-wise differences between rendered and input images via gradient descent. Similarly, iComMa~\cite{sun_icomma_2024} applies gradient-based optimization with a 3DGS model, employing a hybrid loss function combining pixel-wise differences and keypoint-based matching to improve robustness against poor initialization. Both iNeRF and iComMa rely on the Adam optimizer~\cite{adam}, with iComMa demonstrating faster convergence due to the efficient rendering capability of 3DGS. However, even with this improvement, iComMa still requires approximately one second per optimization on an NVIDIA RTX6000 Ada GPU. While our method adopts a similar gradient-based optimization framework as these prior approaches, our primary objective differs fundamentally: we specifically target fast and effective image compression for real-time transmission during inspection missions. To achieve this, our approach leverages temporal continuity to efficiently optimize the latent representation.

\section{Methods}
\label{sec:methods}

\subsection{Overview of the Compression Framework}
We propose an NVS prior-based image compression framework, termed \emph{NVSPrior}, which exploits prior scene knowledge to compress images captured by an ROV during underwater inspection missions. The motivation for this approach is that in many inspection scenarios, the environment is largely static and predictable due to repeated surveys in structured settings. This allows a trained NVS model to reconstruct most of the visual content from a compact latent representation, such as the camera pose and, optionally, transient embeddings~\cite{martin-brualla_nerf_2021}. The more informative the prior, the better the compression—giving NVS a distinct advantage through its ability to generate photorealistic outputs.

The process begins with a mapping run, during which the ROV collects camera images to characterize the environment. These images are used to train a scene-specific NVS model (e.g., 3DGS), and identical copies of the trained model are stored on both the ROV and the operator side to ensure consistent rendering and reconstruction.

During subsequent inspection runs, the ROV estimates the optimal latent representation that allows the NVS model to render an image closely matching the current camera image. To account for scene elements not present during training (such as novel objects or lighting changes), the difference between the camera image and the NVS-rendered image is computed, denoted as $I_{\text{diff}}$, and compressed using a classical image codec.

\begin{figure}
\centering
\includegraphics[width=\linewidth]{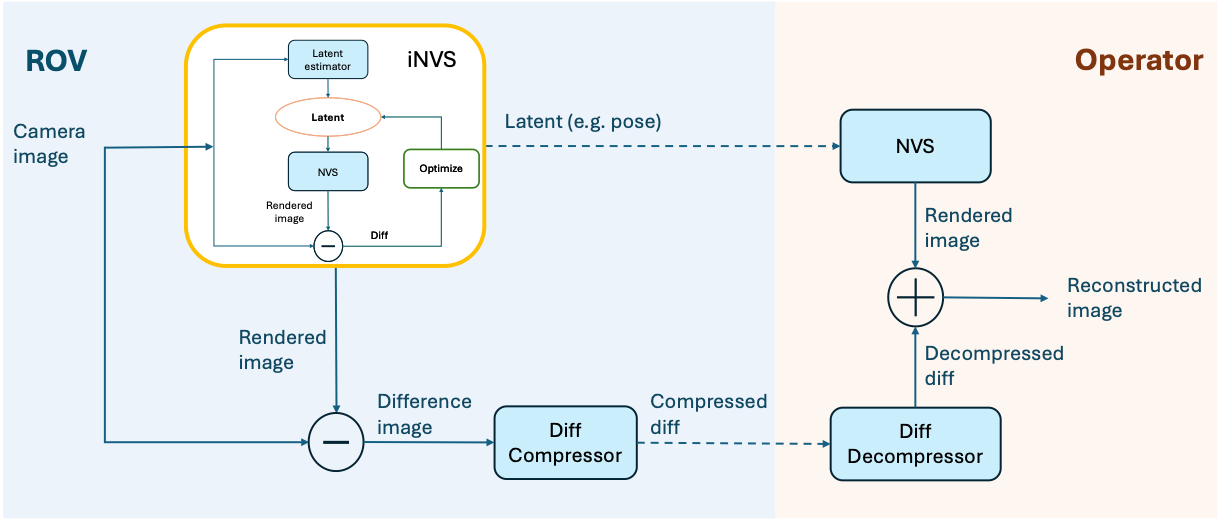}
\caption{Schematic overview of the proposed NVS-based image compression pipeline. Onboard the ROV (blue shaded area), the system encodes a compact latent representation and computes the difference image between the captured frame and the NVS-rendered view. Both the optimized latent and the compressed difference image are transmitted to the operator side (orange shaded area), where the original image is reconstructed using the shared NVS model. This framework enables efficient, high-fidelity image transmission over bandwidth-limited underwater acoustic links.}
\label{fig:proposed-overview}
\end{figure}

Both the optimized latent representation and the compressed $I_{\text{diff}}$ are transmitted to the operator. On the surface, the NVS model uses the received latent representation to generate a rendered image, and the decompressed $I_{\text{diff}}$ is added to reconstruct an estimate of the original camera view. Since most of the scene can be reconstructed from the prior, $I_{\text{diff}}$ is typically highly compressible, resulting in a significant reduction in transmission size. This process is illustrated in Fig.~\ref{fig:proposed-overview}.

This strategy leverages prior knowledge for highly efficient image transmission, making real-time, high-fidelity visual feedback feasible even over bandwidth-constrained underwater acoustic links.

\begin{figure*}[h]
  \centering
  \includegraphics[width=\textwidth]{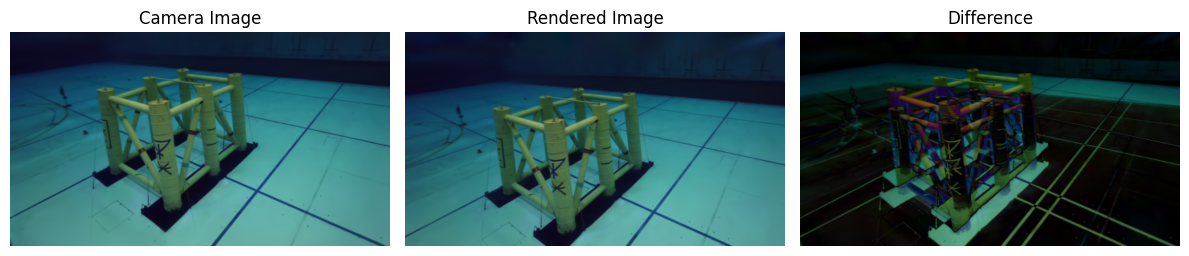}
  \caption{Effect of a 5\textdegree~rotation error on the rendered image. The left image is the camera image, the middle image is the image rendered at the latent representation rotated by 5\textdegree~about the x-axis, and the right image is the difference between the two images.}
  \label{fig:perturbation-effect}
  \end{figure*}

The remaining challenge in this approach lies in its sensitivity to the accuracy of the latent representation—an offset of just a few pixels between the rendered and captured images can significantly increase the size of the difference image. While existing underwater image-based pose estimators~\cite{peng_regressing_2022, peng_improved_2024, peng_pose_2024} can achieve accuracy sufficient for navigation, these estimated latent representations are often not accurate enough for image compression. Depending on the distance from the ROV to the object of interest, a small error in the latent representation can result in a large image difference between the camera image and the rendered image as shown in Fig.~\ref{fig:perturbation-effect}. While prior work~\cite{rajat_oceans_2024} used affine warping to align the rendered and real images, this solution assumes a 2D scene geometry and introduces visual artifacts, which increases the entropy of $I_{\text{diff}}$ and complicates decoding.

To overcome this, we introduce inverse NVS (iNVS), a gradient-based optimization strategy that searches for the latent representation that minimizes the difference between the NVS-rendered image and the actual camera image. This yields a more accurate reconstruction and significantly reduces the size of the difference image to be transmitted.

\subsection{iNVS}
\label{subsec:invs}
\begin{figure}[!h]
  \centering
  \includegraphics[width=0.5\columnwidth]{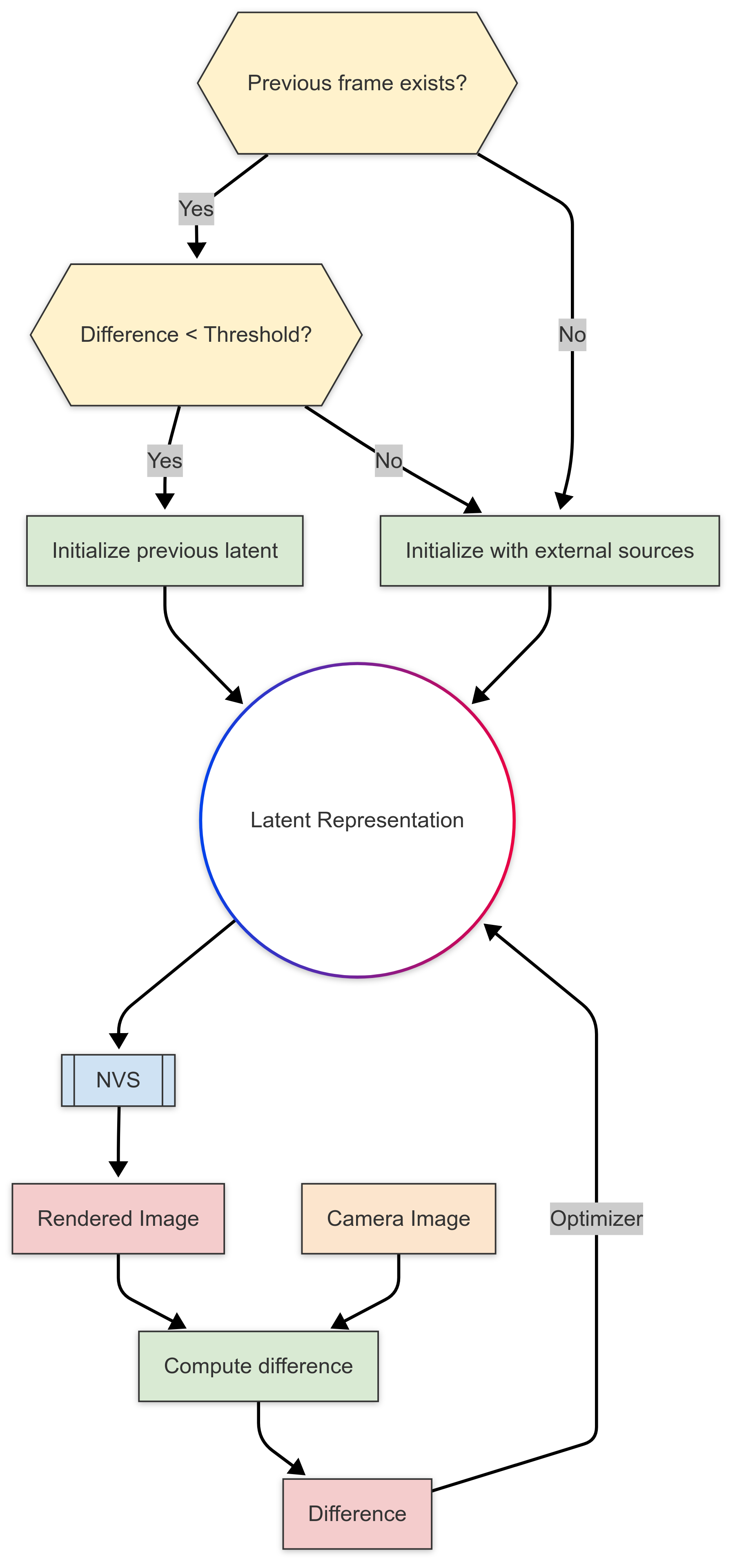}
  \caption{Flow diagram of the iNVS optimization process. The system determines whether to initialize the latent representation using the previous frame or external sources based on availability and difference threshold. The initialized representation is iteratively refined by minimizing the difference between the rendered and camera images using a pretrained NVS model. The resulting optimized latent representation is transmitted along with the residual image as a compressed representation of the image, enabling reconstruction at the topside.}
  \label{fig:flow}
  \end{figure}

iNVS aims to rapidly estimate the latent representation that minimizes the difference between a real camera image and the image rendered by a trained 3DGS model. The steps involved in iNVS are detailed in Fig.~\ref{fig:flow},and its key components—initialization strategy, optimization algorithm, and objective function—are discussed below.

% \begin{algorithm}[ht]
% \caption{iNVS: Inverse Novel View Synthesis for Latent Optimization}
% \label{alg:invs}
% \begin{algorithmic}[1]
% \Require Camera image $I_{\text{cam}}$, previous latent $z_{\text{prev}}$ (optional), external initialization $z_{\text{ext}}$, NVS model $M$, threshold $\tau$
% \Ensure Optimized latent representation $z^*$

% \If{$z_{\text{prev}}$ exists and $\text{MSE}(I_{\text{cam}}, M(z_{\text{prev}})) < \tau$}
%     \State $z_0 \gets z_{\text{prev}}$
% \Else
%     \State $z_0 \gets z_{\text{ext}}$ \Comment{Fallback to external source (e.g., sensor or estimator)}
% \EndIf

% \State Initialize $z \gets z_0$
% \Repeat
%     \State $I_{\text{render}} \gets M(z)$
%     \State $\mathcal{L} \gets \text{MSE}(I_{\text{cam}}, I_{\text{render}})$
%     \State Update $z$ to minimize $\mathcal{L}$ using a gradient-based optimizer
% \Until{convergence criteria are met}
% \State \Return $z^* \gets z$
% \end{algorithmic}
% \end{algorithm}

% \begin{figure}[!h]
%   \centering
%   \includegraphics[width=\columnwidth]{figures/flowchart.png}
%   \caption{Flow of iNVS.}
%   \label{fig:flow}
% \end{figure}
The first step is to initialize the latent representation that we aim to estimate. An effective initialization is crucial for rapid convergence of optimization algorithms. In inspection missions where the vehicle moves slowly and steadily, there are minimal changes in the latent representation between consecutive frames. Assuming such a scenario, we use the optimized latent representation from the previous frame as an initialization point to estimate the latent representation in the current frame, provided that a ``good'' previous frame exists. This approach leverages the small inter-frame variations in the latent representation, enabling faster convergence due to the proximity of the initial estimate to the true latent representation. By utilizing the optimized latent representation from the previous frame whenever possible, we mitigate the issues associated with sensor drift, biases, and estimator noise, providing a more accurate starting point for optimization.

To determine whether a previous frame is ``good'', we compare the rendered image at that latent representation with the current camera image. If their difference falls below a predefined threshold, the estimate is reused. When a good previous frame is unavailable, such as at the start of a mission, or when the difference exceeds the threshold, alternative initialization sources are employed. These may include measurements from vehicle sensors or estimates from learned latent estimators.

The optimization step in iNVS refines the initial latent representation by minimizing a differentiable image similarity loss between the rendered and observed images. Since the number of parameters to be optimized is small (typically a 6-DoF pose), both deterministic and stochastic optimization methods can be considered. In our implementation, we explore both a quasi-Newton method and a stochastic gradient-based method, as described in Section~\ref{subsec:implementation}.

The objective function quantifies the discrepancy between the rendered image and the camera image. A commonly used choice is the mean squared error (MSE), defined as:

\begin{equation}
\label{eq:mse}
    L_{\text{mse}} = \frac{1}{N}\sum_{i=1}^{N} \lVert I_{\text{camera}}(i) - I_{\text{rendered}}(i) \rVert_2^2
\end{equation}
where $I_{\text{camera}}(i)$ and $I_{\text{rendered}}(i)$ are the RGB vectors (e.g., 3-dimensional) at the $i^\text{th}$ pixel for the camera image and the rendered+affine-transformed image, respectively. $\lVert \cdot \rVert_2$ denotes the Euclidean (L2) norm across the color channels, and $N$ is the total number of pixels.

Alternative loss functions, such as the keypoint-based matching loss, may improve robustness under poor initialization~\cite{sun_icomma_2024}. These are evaluated in Section~\ref{subsec:config-choices}.

\section{Experiments \& Results}
\label{sec:experiments}

\subsection{Dataset}
\subsubsection{Controlled Environment}
\begin{figure*}[t!]
  \centering
  \begin{subfigure}{0.31\textwidth}
      \centering
      \includegraphics[width=\textwidth]{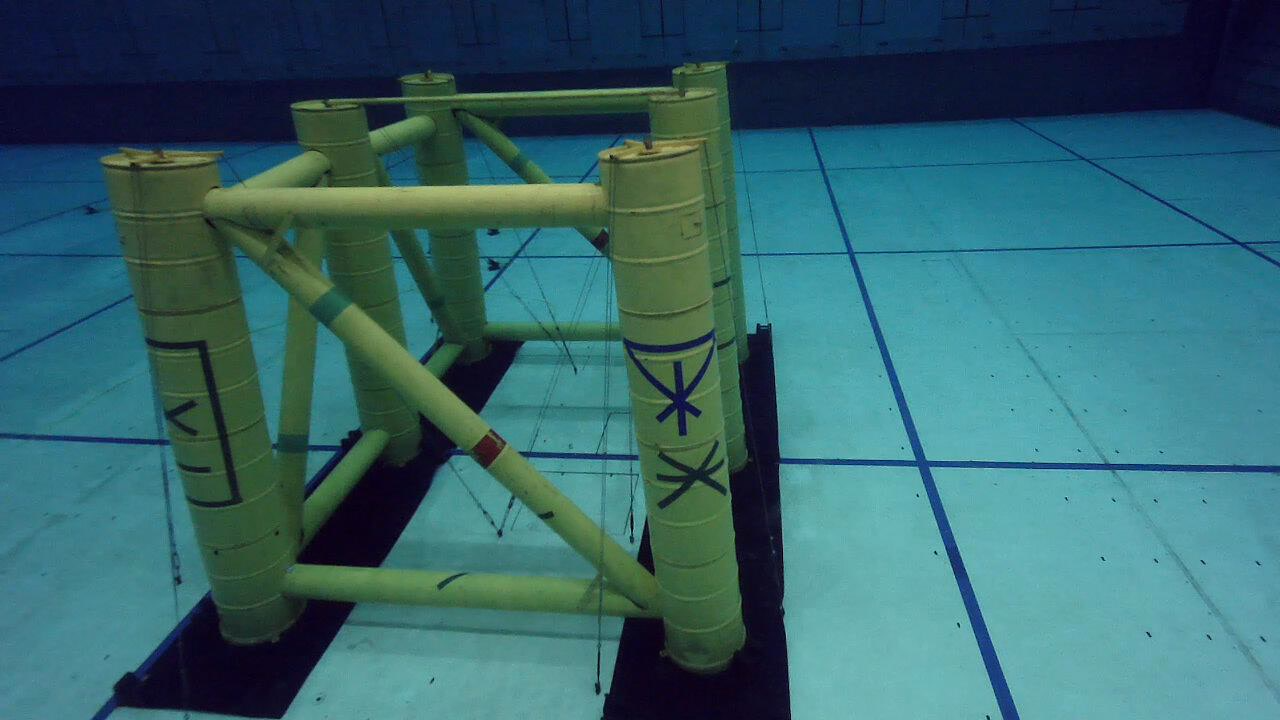}
     \caption{Example image from M1.}
  \end{subfigure}\hfill
  \begin{subfigure}{0.31\textwidth}
    \centering
    \includegraphics[width=\textwidth]{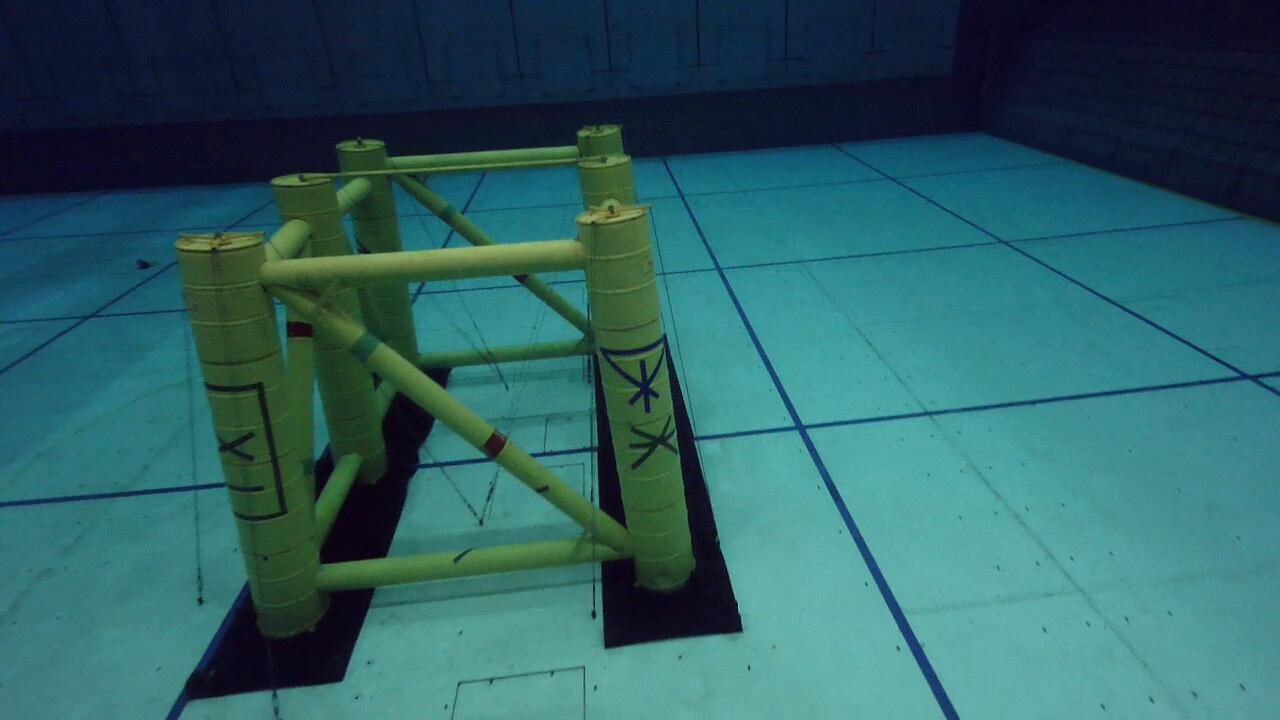}
     \caption{Example image from T1.}
  \end{subfigure}\hfill
  \begin{subfigure}{0.31\textwidth}
    \centering
    \includegraphics[width=\textwidth]{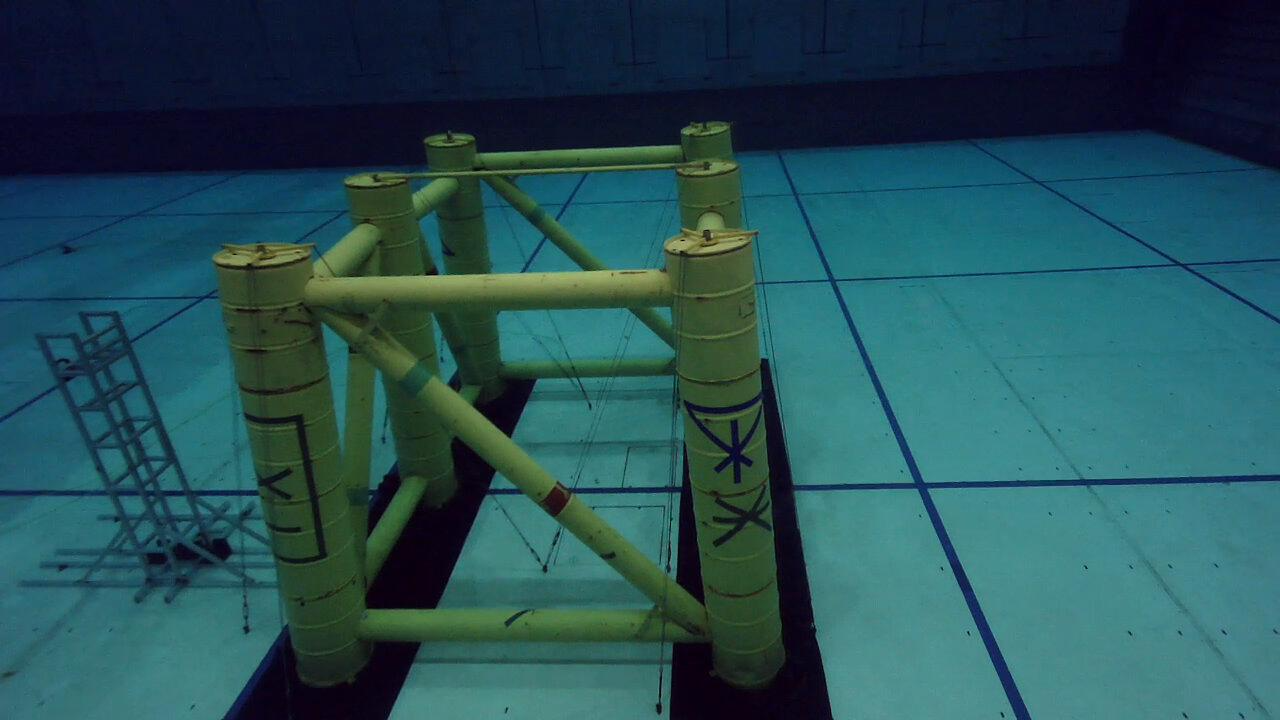}
     \caption{Example image from T2.} 
     \label{fig:new_struct}
  \end{subfigure}
  \caption{Example images from the controlled environment dataset. Panel (c) features T2 where an additional metallic structure placed next to the original structure to test the robustness of our technique.}

  \label{fig:tcoms_struct}
\end{figure*}

To systematically evaluate algorithmic choices for iNVS and assess our image compression technique in a controlled setting, we conducted experiments at the Technology Center for Offshore and Marine, Singapore (TCOMS)\cite{TCOMS}. The facility features an indoor artificial ocean basin measuring 60m~$\times$48m~$\times$12m. As illustrated in Fig.~\ref{fig:tcoms_struct}, we placed a yellow structure in the basin consisting of six piles interconnected by metallic pipes, with each pile comprising three metallic oil barrels. The overall dimensions of the structure were approximately 3.9~m $\times$ 4.6~m $\times$ 3.0~m.

We conducted three survey runs within this environment, where the ROV surveyed the structure using a similar lawnmower trajectory in each trial. We refer to these trials as the mapping run (M1), test run 1 (T1), and test run 2 (T2). In M1, the ROV collected data for training the 3DGS model which forms the prior. In T1 and T2, we evaluate the performance of our proposed image compression technique using the model trained based on M1. As shown in Fig.~\ref{fig:new_struct}, during T2, a new metallic structure was placed next to the existing one to test the robustness of our technique towards novel objects in the scene.

We selected 1,422 images from M1 for training, and sampled 1,000 consecutive images from each of T1 and T2 for evaluation. The images collected during the trials were color images with a resolution of 1280 $\times$ 720 pixels, captured at 3 frames per second. We downscaled these images to 320$\times$180 pixels for transmission purposes, and the image compression is tested at this image resolution.

\subsubsection{Real-world Dataset}

To evaluate the robustness and feasibility of the proposed method under real-world underwater conditions, we conduct experiments on two publicly available datasets.

We first evaluate on the IUI3-RedSea scene from the SeaThru-NeRF dataset~\cite{seathru}. As the most widely used benchmark for underwater novel view synthesis, this dataset consists of forward-facing imagery of a small coral site captured under natural lighting conditions. Although the scene contains dynamic elements such as moving fish and exhibits underwater effects including backscattering and wavelength-dependent color attenuation, the images are relatively clear compared to deep-water inspection data. The dataset contains only 29 images in total; consequently, all images are used for both training and evaluation.

\begin{figure*}[t!]
  \centering
  \begin{subfigure}{0.49\textwidth}
      \centering
      \includegraphics[width=\textwidth]{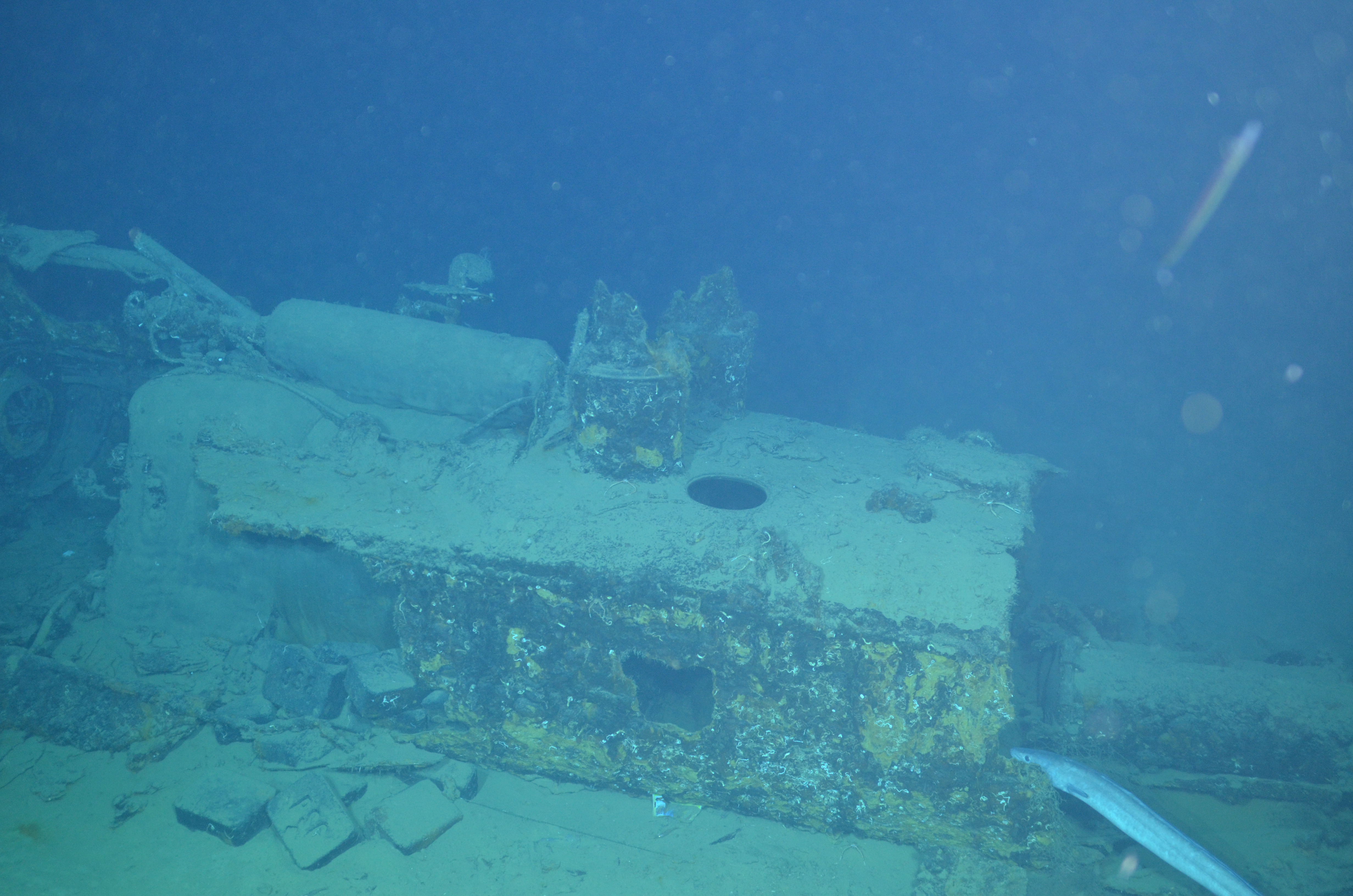}
      \caption{Raw image shows severe color attenuation and low contrast.}
  \end{subfigure}\hfill
  \begin{subfigure}{0.49\textwidth}
    \centering
    \includegraphics[width=\textwidth]{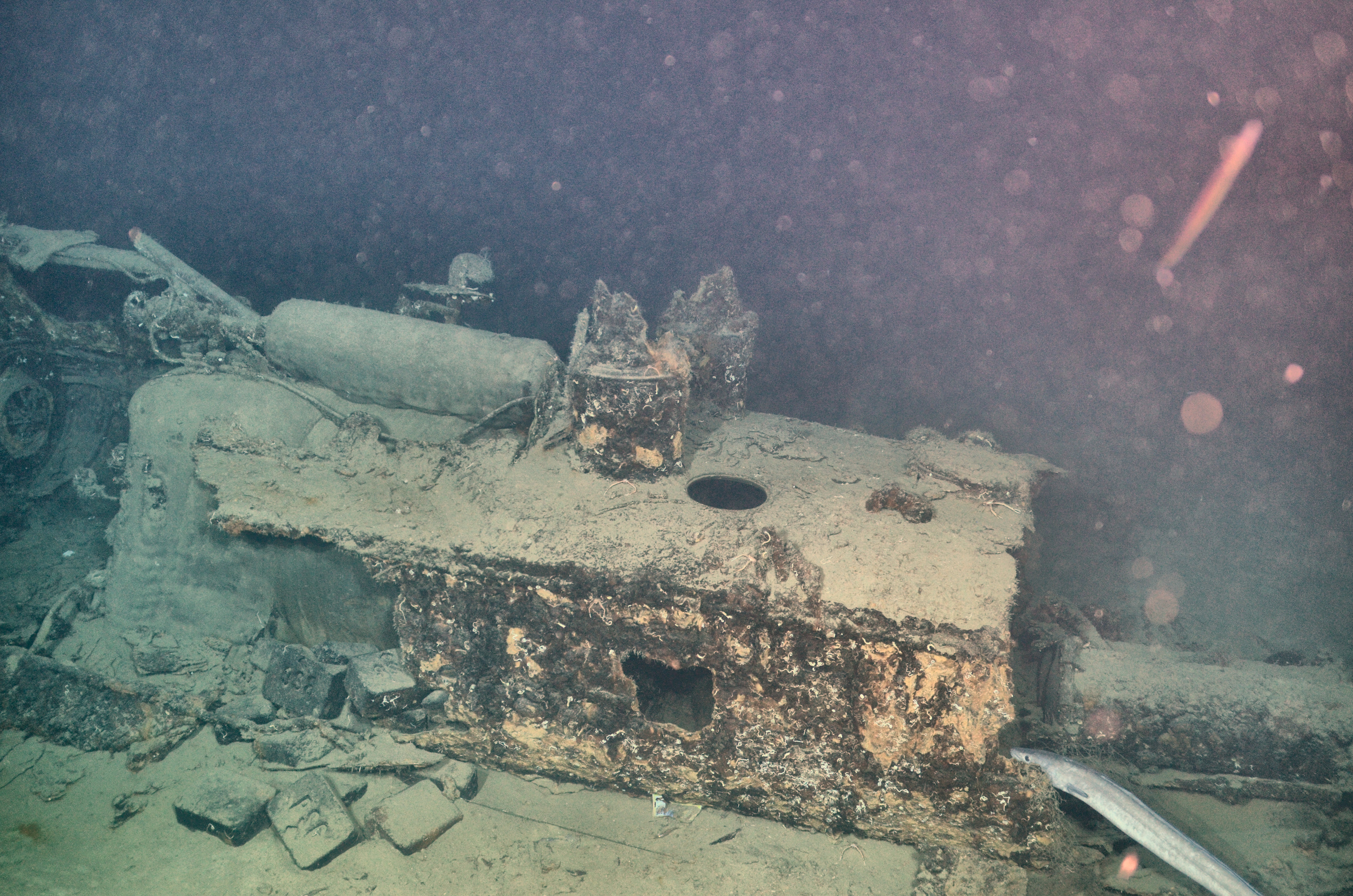}
    \caption{The corresponding image after preprocessing.}
  \end{subfigure}\hfill
  \caption{Example images from the Torpedo Boat Wreck dataset~\cite{torpedo}. This dataset is characterized by challenging imaging conditions, including inconsistent lighting, significant marine snow, and dynamic content such as moving fish.}
  \label{fig:torpedo}
\end{figure*}

We further evaluate on the Torpedo Boat Wreck dataset~\cite{torpedo}, which represents a more challenging deep-water inspection scenario. The dataset was collected during a survey around a submerged wreck measuring approximately 20~m in length, 3~m in width, and rising about 2~m above the surrounding seafloor, at a depth of roughly 476~m. The survey provides near-360° visual coverage of the wreck. Due to the use of artificial illumination at depth, the raw images exhibit strong color attenuation, low contrast, and significant visual degradation. We apply Gray-world white balancing~\cite{grayworld} and contrast-limited adaptive histogram equalization (CLAHE)~\cite{clahe} as preprocessing steps to partially mitigate these effects. Nevertheless, the dataset remains highly challenging due to persistent marine snow (shown in Fig.~\ref{fig:torpedo}), moving fish, and large viewpoint changes caused by an inconsistent frame rate. The dataset contains 442 sequential images, of which 147 are used for evaluation and the remainder for training.

For both datasets, all images are downscaled to a resolution of 320~$\times$~180 pixels for consistency in performance evaluation. As both datasets are limited in scale compared to typical operational inspection missions, this data scarcity limits the training of robust scene priors and pose estimators, thereby providing a stringent stress test for the proposed method.

\begin{figure*}[t!]
  \centering
  \begin{subfigure}{0.49\textwidth}
      \centering
      \includegraphics[width=\textwidth]{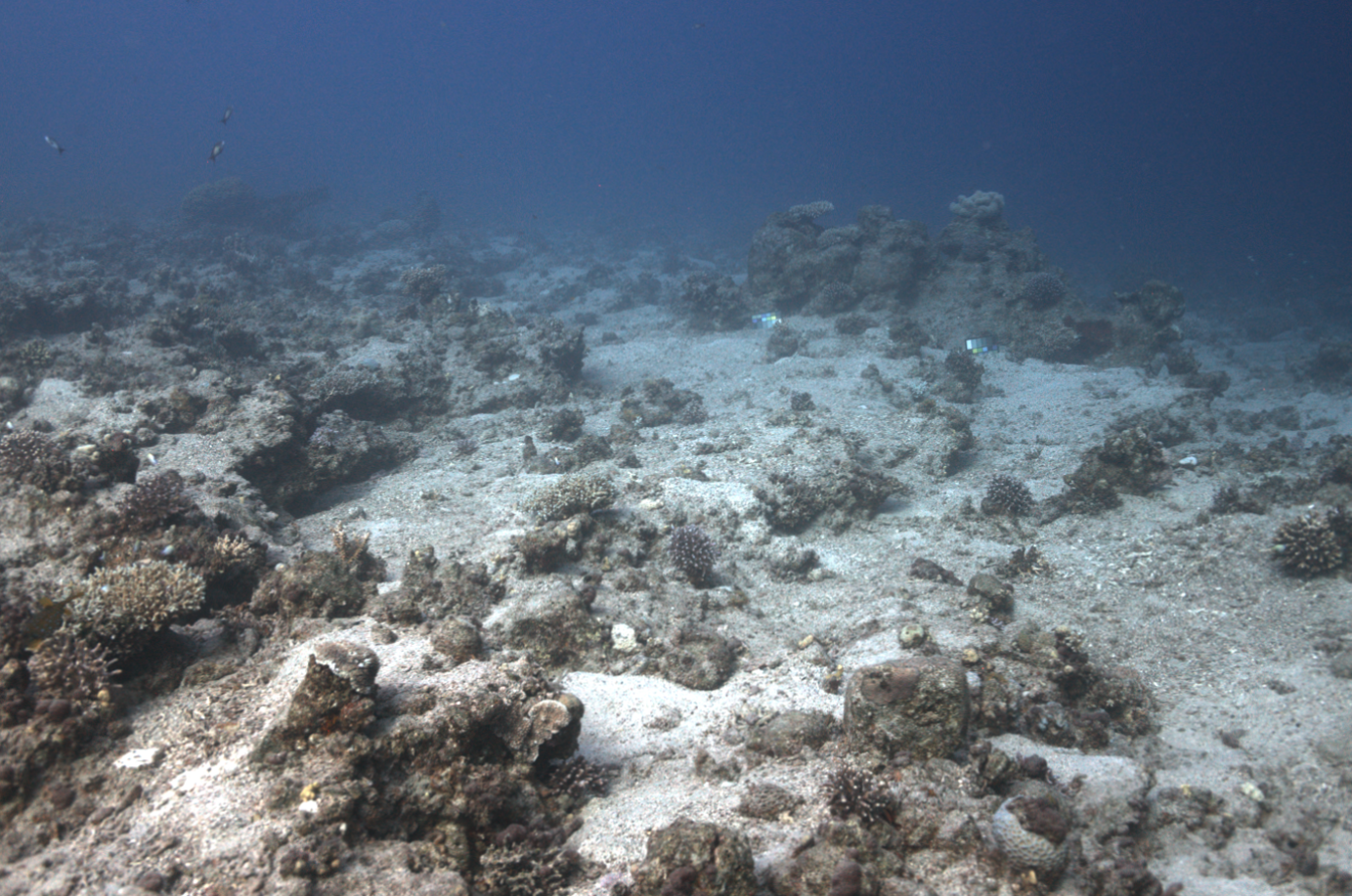}
  \end{subfigure}\hfill
  \begin{subfigure}{0.49\textwidth}
    \centering
    \includegraphics[width=\textwidth]{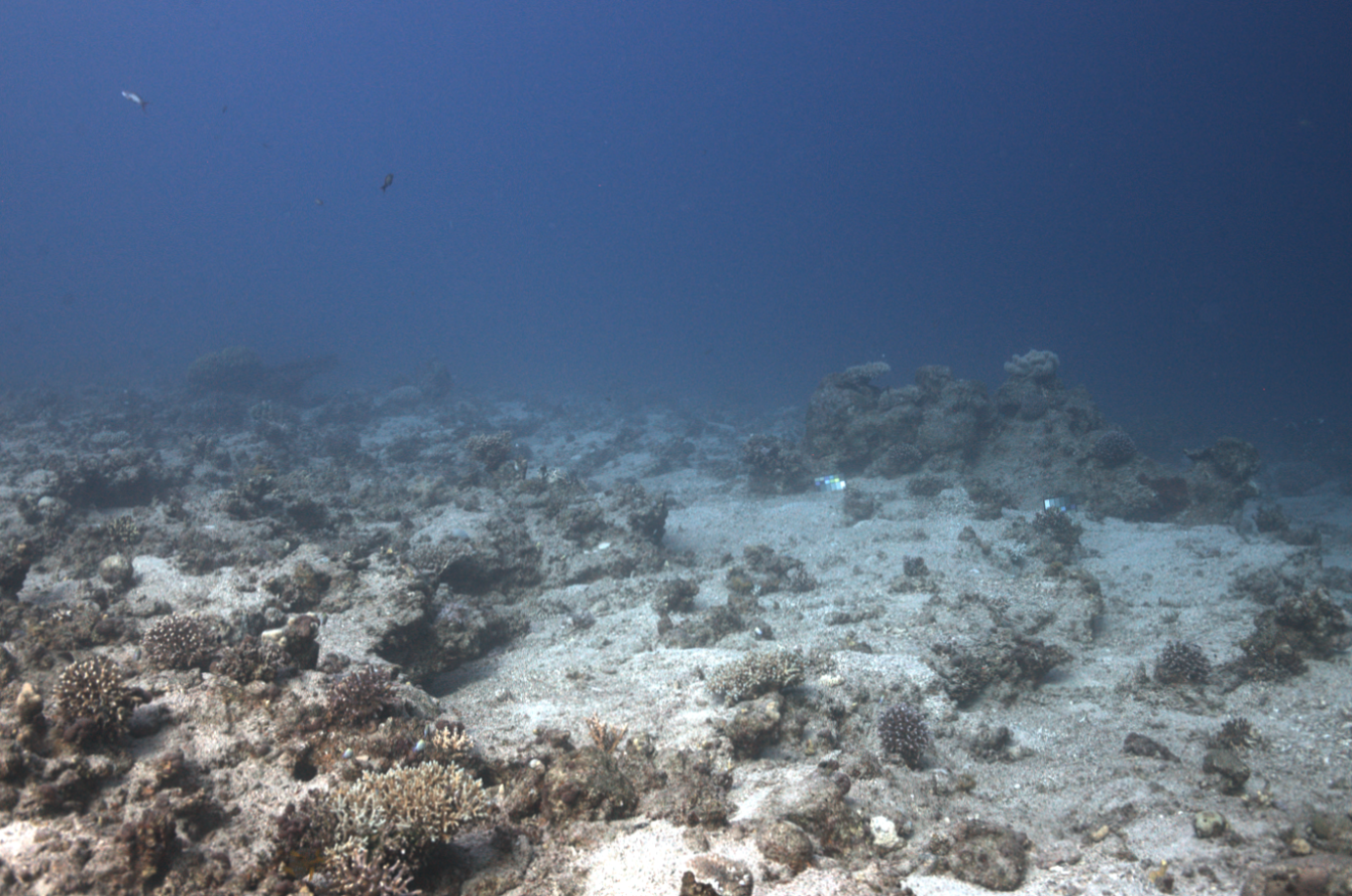}
  \end{subfigure}\hfill
  \caption{Example images from the IUI3-RedSea scene of the SeaThru-NeRF dataset~\cite{seathru}. The images illustrate typical underwater challenges including backscattering, color attenuation, and dynamic elements such as the moving fish visible at the top left corners.}
  \label{fig:redsea}
\end{figure*}

\subsection{Implementation Details}
\label{subsec:implementation}

\subsubsection{iNVS Configuration}
\paragraph{Underlying Scene Representation.}
We adopt 3DGS due to its high rendering speed, which enables real-time optimization. To train the 3DGS model, we use camera images from the initial survey run (the ``mapping run'') and employ the Splatfacto pipeline~\cite{splatfacto} provided by the Nerfstudio framework~\cite{nerfstudio}.

\paragraph{Pose Initialization}
For the controlled dataset, we use PoseLSTM, a neural network trained to estimate 6-DoF camera poses from RGB images. It comprises a pretrained ResNet-50~\cite{he_deep_2016} based feature extractor and a bidirectional LSTM for dimensionality reduction. Ground-truth poses for training are generated via COLMAP~\cite{schonberger_pixelwise_2016}, an open-source structure-from-motion tool. These are matched with corresponding images from the mapping run to form the training dataset.

For the real-world dataset, the number of available images is insufficient to train a dedicated pose estimator. Instead, we compute camera poses using COLMAP and simulate pose estimation noise by perturbing the poses with uniformly distributed noise: we vary translations within the range of [$-0.11$~m, $0.11$~m], and rotations within [-3\textdegree, 3\textdegree]. These bounds reflect the expected accuracy of typical underwater pose estimators~\cite{peng_pose_2024}. We use the perturbed poses as initializations for iNVS in lieu of a trained estimator.

To determine whether the estimated latent representation is suitable for initializing the next frame, we evaluate the MSE between the normalized camera image and the normalized NVS-rendered image at the optimized latent. We set a threshold of $1 \times 10^{-3}$ for the controlled dataset and a more relaxed threshold of $1 \times 10^{-2}$ for the real-world dataset to account for increased visual variability and noise.

\paragraph{Objective Function}
We evaluate two loss functions for pose refinement. The first is the MSE between the rendered and camera images, defined in~\eqref{eq:mse}, which provides a direct pixel-level comparison and is computationally efficient. The second is a keypoint-based matching loss~\eqref{eq:match} introduced by iComMa~\cite{sun_icomma_2024}, which uses LoFTR~\cite{sun_loftr_2021} to extract and match keypoints across the two images. This approach aims to improve robustness under poor initialization by focusing on structurally meaningful features rather than pixel-level alignment. While LoFTR can produce a large set of correspondences, using all matches incurs substantial computational overhead~\cite{sun_loftr_2021}. To meet our runtime budget, we compute the matching loss using only the top-$K$ correspondences ranked by LoFTR confidence (we use $K{=}20$), which also reduces the impact of ambiguous matches in repetitive-texture regions commonly observed in underwater scenes. The robustness of the objective functions is evaluated in Section~\ref{subsec:config-choices}.

The matching loss is defined as:
\begin{equation}\label{eq:match}
    L_{\text{match}} = \frac{1}{M}\sum_{i=1}^{M} \left\| \mathbf{k}_{\text{camera}}(i) - \mathbf{k}_{\text{rendered}}(i) \right\|^2
\end{equation}
where \( \mathbf{k}_{\text{camera}}(i) \) and \( \mathbf{k}_{\text{rendered}}(i) \) denote the coordinates of the \(i^\text{th}\) matched keypoint pair in the camera image and the rendered image, respectively, as determined by LoFTR. Each keypoint \( \mathbf{k}(i) \in \mathbb{R}^2 \) is a 2D coordinate representing a salient point in the image. The loss computes the mean squared Euclidean distance between corresponding matched keypoints across the two images, serving as a measure of geometric alignment between the rendered image and the observed camera view.

\paragraph{Optimization Method}
For refining the latent representation, we evaluate both deterministic and stochastic optimization methods. As the deterministic optimizer, we select the Broyden-Fletcher-Goldfarb-Shanno (BFGS) algorithm~\cite{bfgs_broyden, bfgs_fletcher, bfgs_goldfarb, bfgs_shanno}, which is a quasi-Newton method that approximates the Hessian matrix using gradient evaluations and iteratively updates parameters to converge toward a local minimum. This choice is motivated by the fact that our search involves a small number of parameters (e.g., 6-DoF pose), and assuming that iNVS starts with a reasonably good initialization. In this case, the initialization is often near the global minimum, allowing a deterministic optimizer like BFGS to converge quickly and reliably. The BFGS algorithm is implemented using the pytorch-minimize package~\cite{torchmin}. We set the gradient tolerance to $1 \times 10^{-5}$ and the parameter tolerance to $1 \times 10^{-6}$ to ensure convergence.

We also implement Adam~\cite{adam} as a stochastic optimizer using PyTorch, using an initial learning rate of $1 \times 10^{-3}$ and halving it if no loss improvement is observed over three epochs. These configurations are compared in Section~\ref{subsec:config-choices}.

\paragraph{Difference Image Compression}
We compress the difference image $I_{\text{diff}}$ between the camera and rendered views using WebP or JPEG-XL, chosen due to their support for fast encoding and decoding.

\subsubsection{Benchmarking Methods}
We compare against two families of baselines: (i) classical codecs—WebP and JPEG-XL, and (ii) learned codecs implemented with CompressAI~\cite{compressai}, namely, the Mean \& Scale Hyperprior method~\cite{mbt18} and MLIC++~\cite{mlic}. For classical codecs, we compress each RGB frame after resizing to 320×180 pixels.

For the learned baselines, we train the model on resized RGB frames with the standard rate–distortion objective
\[
\mathcal{L} = \mathcal{R} + \lambda\,255^{2}\,\mathcal{D},
\]
where \(\mathcal{R}\) denotes bitrate and \(\mathcal{D}\) is the MSE computed on images normalized to \([0,1]\); the \(255^{2}\) factor follows CompressAI’s convention. A
larger \(\lambda\) emphasizes minimization of distortion, yielding higher reconstruction quality.
We report the highest-quality operating points using the largest \(\lambda\) provided by each implementation.

We also benchmark against NVSPrior + Affine, an adaptation of the approach proposed by Mishra et al.~\cite{rajat_oceans_2024}, which uses a learned affine transformation to align the rendered image with the camera image. In this method, we first estimate the latent representation using PoseLSTM and render the image using the 3DGS model. We then compute affine transformation parameters to align the rendered image with the camera view. The resulting difference image is compressed using either WebP or JPEG-XL. This approach serves as a baseline to assess the benefits of our proposed iNVS method, which directly refines the latent representation to minimize the difference image without relying on affine warping.

All experiments are conducted on a workstation equipped with an AMD Ryzen™ 9 7950X CPU (32 threads), 128 GB RAM, and an NVIDIA RTX6000 Ada GPU. The primary software used includes Nerfstudio 1.1.4, along with PyTorch 2.1.2.

\subsection{Ablation studies}
\label{subsec:config-choices}
In this subsection, we conduct ablation studies on the proposed iNVS, evaluating different loss functions and optimization algorithms to assess their impact on compression performance and robustness to pose initialization errors. Unlike classical or learned codecs, the NVSPrior framework is directly influenced by such errors, making this analysis essential for identifying the most effective configuration.
\subsubsection{Performance and robustness using MSE and Matching Loss}
In this subsection, we study the performance of iNVS with different objective functions, and how it varies with the error in the initial latent representation. We introduce perturbations to the latent representation to simulate initialization errors that occur during actual survey runs. For each perturbation, we randomly select either a translation or rotation axis and sample a perturbation value from a uniform distribution within specified ranges. Translation perturbations are sampled uniformly within [-1.58~m,1.58~m], and rotation perturbations are sampled uniformly within [-40°, 40°]. Using the perturbed latent representation as the initialization, we optimize the latent representation using either MSE loss or matching loss until convergence.

\begin{figure*}[!h]
  \centering
  \begin{subfigure}{\textwidth}
      \centering
      \includegraphics[width=\textwidth]{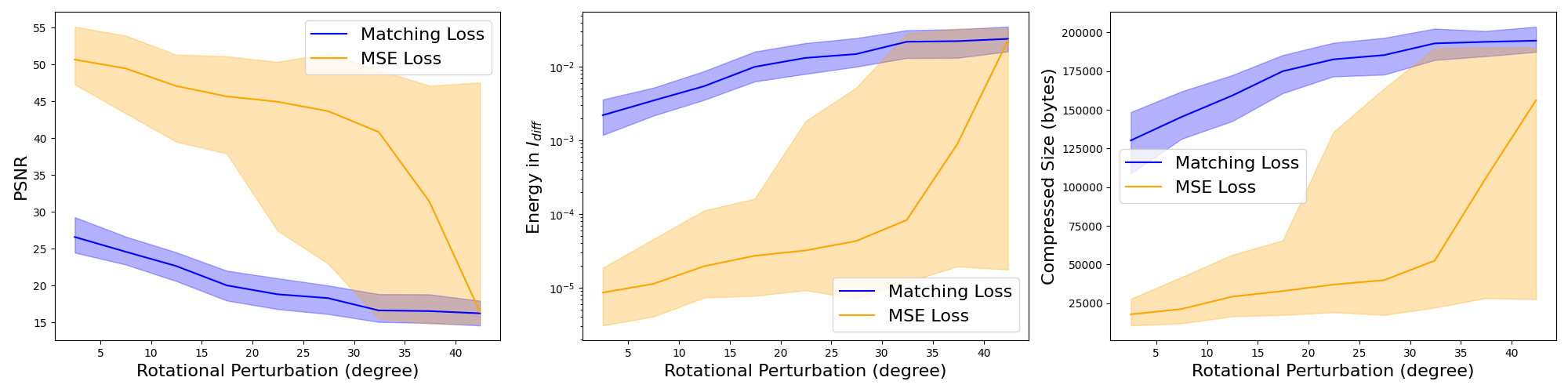}
     \caption{Performance variation under rotational perturbations. }
  \end{subfigure}\hfill
  \begin{subfigure}{\textwidth}
    \centering
    \includegraphics[width=\textwidth]{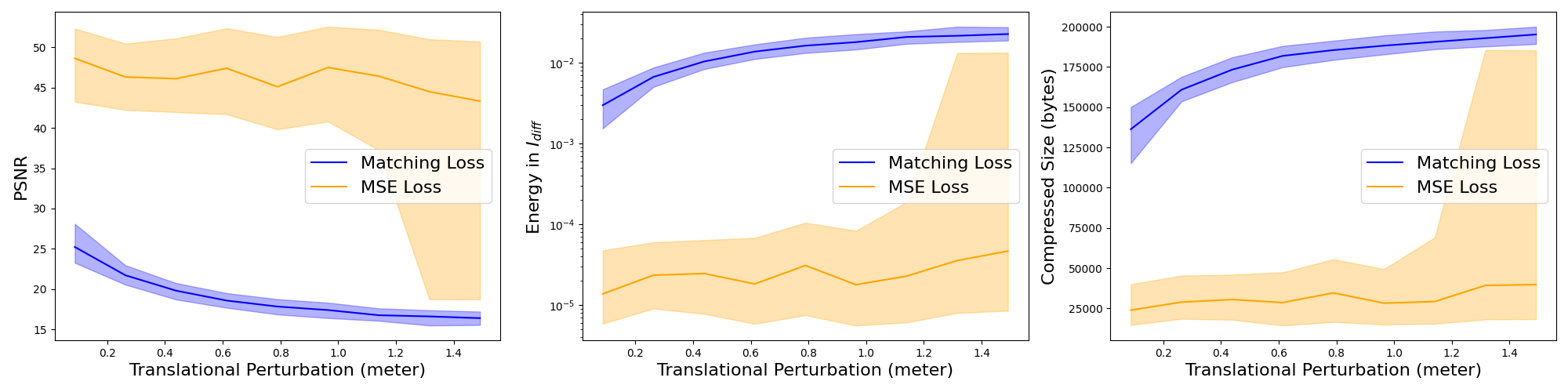}
     \caption{Performance variation under translational perturbations. }
  \end{subfigure}\hfill
  \caption{Comparison of performance using MSE loss and Matching Loss as objective functions across different levels of initialization perturbation. In each ribbon plot, the solid line indicates the median value, while the shaded region denotes the interquartile range across samples. Metrics include PSNR, energy of the difference image, and compressed size.
  }
  \label{fig:loss_funcs}
\end{figure*}

After optimization, we compute the Peak Signal-to-Noise Ratio (PSNR) of the rendered images compared to the camera images in dB, energy in $I_{\text{diff}}$ which is equivalent to the MSE~\eqref{eq:mse}, and the number of bytes of compressed $I_{\text{diff}}$ between the camera image and rendered image at the optimized latent representation. These metrics jointly capture both visual fidelity and communication cost: PSNR reflects reconstruction quality, energy measures the magnitude of residual errors, and compressed size relates directly to the data rate, which is the ultimate focus of this work. We also record the number of iterations required for convergence. This process is repeated on the 1,422 images from M1. The results are shown in Fig.~\ref{fig:loss_funcs}. 
The PSNR is defined as:
\begin{equation}
\label{eq:psnr}
\text{PSNR} = 10 \cdot \log_{10} \left( \frac{1}{L_{\text{mse}}} \right)
\end{equation}
where $L_{\text{mse}}$ is defined in~\eqref{eq:mse}.

The model trained using MSE loss outperforms that using matching loss across all metrics and perturbation values in terms of median performance, despite exhibiting higher variance. Overall, using MSE loss achieves a rendered image with higher PSNR and thus a more compressed $I_{\text{diff}}$ than using matching loss. The larger variability in performance observed when using MSE loss is likely due to its sensitivity to initialization and the presence of more local minima in its loss landscape. Moreover, MSE loss converges faster than matching loss, even though it requires more iterations. As a result, we select MSE loss as the objective function for iNVS.

\subsubsection{Optimization Methods}
\begin{figure*}[!h]
  \centering
  \begin{subfigure}{\textwidth}
      \centering
      \includegraphics[width=\textwidth]{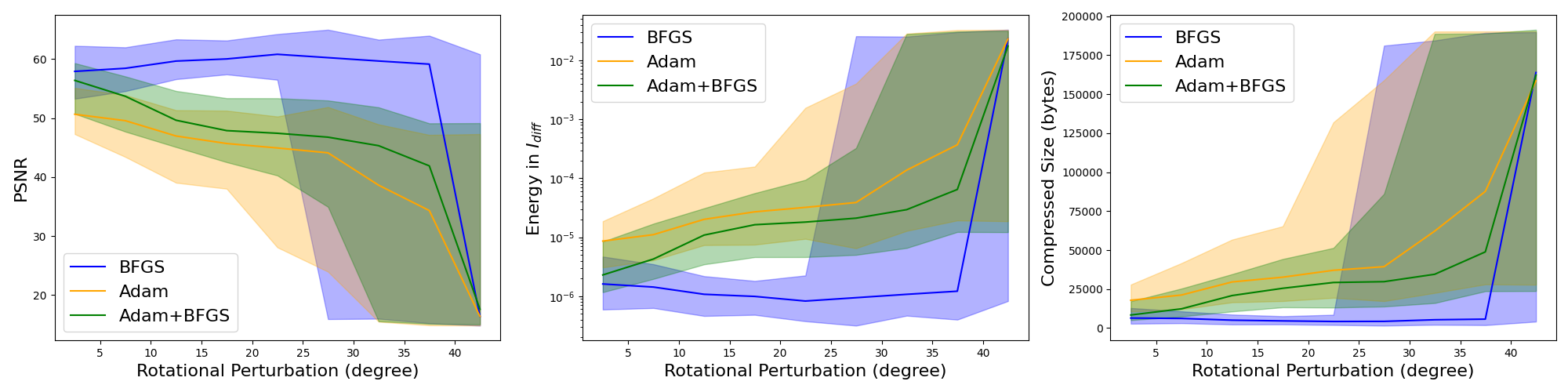}
     \caption{Performance variation under rotational perturbations. }
  \end{subfigure}\hfill
  \begin{subfigure}{\textwidth}
    \centering
    \includegraphics[width=\textwidth]{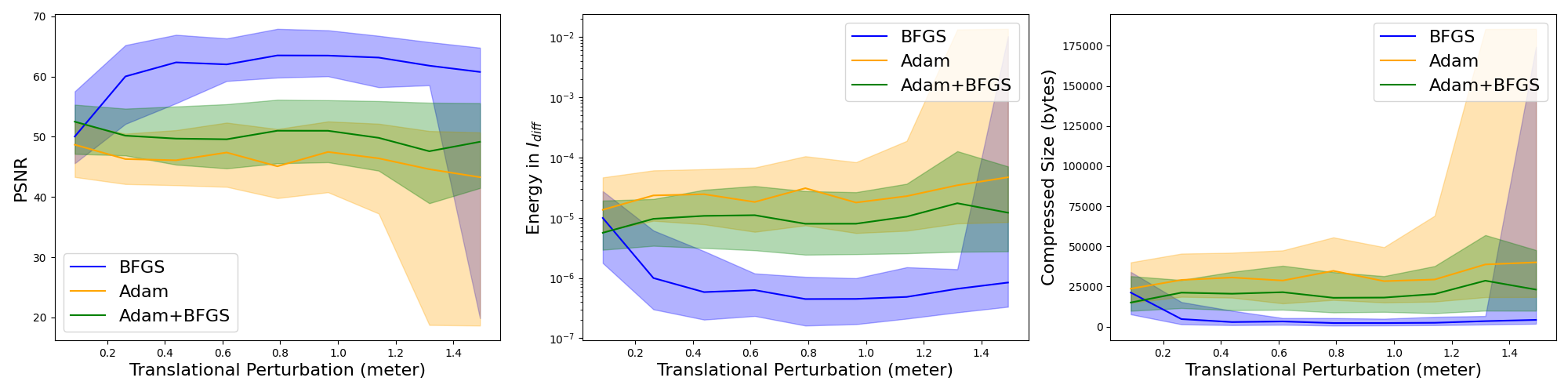}
     \caption{Performance variation under translational perturbations. }
  \end{subfigure}\hfill
  \caption{Comparison of performance using Adam and BFGS as optimizers across different levels of initialization perturbation.}
  \label{fig:optimizers}
\end{figure*}

With the MSE loss as the objective function, we compare the performance of the BFGS and Adam optimization algorithms using a similar approach as described above. Additionally, we investigate the performance of a hybrid method, denoted as Adam+BFGS, where Adam is used as the optimizer until convergence, followed by BFGS for finetuning (a common practice in optimization~\cite{Gamper2024A,Dharanalakota2023Loss-based}). The results, presented in Fig.~\ref{fig:optimizers}, show that both BFGS and Adam+BFGS outperform Adam across all three metrics and perturbation values. Adam+BFGS performs slightly better than BFGS at small translational perturbations; however, overall, BFGS achieves better performance than Adam+BFGS at higher perturbation levels. Adam+BFGS exhibits the smallest variance among all three optimization methods at high perturbations and hence better reliability.

Apart from exhibiting better median performance, BFGS also requires fewer iterations to converge than Adam and Adam+BFGS, and each of its iterations is faster. This is likely because BFGS, a second-order optimizer, leverages curvature information and progresses efficiently even when the initial pose is moderately inaccurate. In contrast, Adam, as a momentum-based first-order method, may converge more slowly or inconsistently due to noisy gradients, particularly in this low-dimensional setting. Although the hybrid Adam+BFGS method is more stable under small perturbations, it adds computational overhead without consistent benefits at larger errors. Therefore, we choose BFGS as the optimization method for iNVS henceforth.

It is important to note that the performance of iNVS degrades rapidly when the perturbations are large (e.g., greater than 1.3~m in translation or 37\textdegree~in rotation). Therefore, a good initialization is crucial for the optimal performance of iNVS.

\subsubsection{Initialization}
We compare the convergence performance of iNVS using two initialization methods: (1) the latent representation estimated from the previous frame and (2) the estimate by PoseLSTM on the current frame. Our results show that initializing with the previous frame's latent representation requires fewer iterations for convergence and is more computationally efficient, as it eliminates the need for neural network inference. 

\subsection{Compression Performance on Controlled Dataset}
\label{subsec:compression-performance-controlled}
We evaluate the compression performance of NVSPrior on the controlled dataset T1 using iNVS, configured with BFGS optimization and MSE loss. For both iNVS and the baselines, we test WebP and JPEG-XL as the codecs for compressing the difference image $I_{\text{diff}}$.

We compare against the following families of baselines: (1) WebP and JPEG-XL applied directly to the original RGB image captured by the ROV camera, (2) Mean \& Scale Hyperprior and MLIC++~\cite{mbt18, mlic}, and (3) NVSPrior + Affine, which uses affine warping.

Our evaluation metrics are:
\begin{enumerate}
    \item \textbf{Transmitted data size} (in bytes): the total size of data required to reconstruct the image. For our method, this includes both the optimized latent representation and the compressed difference image $I_{\text{diff}}$. The size of the latent representation is fixed at 28 bytes, consisting of 7 float32 values (3 for translation and 4 for quaternion rotation). For the classical codecs, we report the size of the compressed image. For the learned compression method, we report the size of the quantized latent representations~\cite{compressai}.
    \item \textbf{Compression ratio}: the ratio between the size of the original uncompressed RGB image (320$\times$180 pixels, 3 bytes per pixel) and the size of the transmitted data.
    \item \textbf{PSNR}: as defined in~\eqref{eq:psnr}, computed between the camera image and reconstructed image, quantifying reconstruction quality.
    \item \textbf{Processing time per frame}: time required to compress and reconstruct an image, reflecting the method’s computational efficiency and real-time feasibility.
\end{enumerate}

We report the results in Table~\ref{tab:t1_results}.

\begin{table}[!h]
  \renewcommand{\arraystretch}{1.3}
  \caption{Quantitative results on the T1 dataset, averaged over 1000 \(320\times180\) RGB images. The columns show Size (of the transmitted data), Ratio (which is the compression ratio), PSNR, and processing time per frame. Arrows show the increasing/decreasing trend of the metric indicating improvement.}
  \label{tab:t1_results}
  \centering
  \begin{tabular}{l|c|c|c|c}
    \hline
    Method & Size (bytes) $\downarrow$ & Ratio $\uparrow$ & PSNR (dB) $\uparrow$ & Time (ms) $\downarrow$\\
    \hline\hline
    WebP & 3544 & 48.76  & 33.30 & $\sim$ 6 \\
    JPEG-XL & 5711 & 30.25 & 33.57 & $\sim$ \textbf{1} \\
    Mean \& Scale Hyperprior~\cite{mbt18} & 14112 & 12.25  & 35.03 & $\sim$22 \\
    MLIC++~\cite{mlic} & 4174 & 41.40 & 31.19 & $\sim$ 129 \\
    NVSPrior+Affine+WebP~\cite{rajat_oceans_2024} & 4164 & 41.50  & 31.85 & $\sim$ 64 \\
    NVSPrior+Affine+JPEG-XL~\cite{rajat_oceans_2024} & 4401 & 39.26 & 31.31 & $\sim$ 59 \\
    NVSPrior+iNVS+WebP (Ours) & \textbf{1219} & \textbf{141.76} & 35.83 & $\sim$ 62 \\
    NVSPrior+iNVS+JPEG-XL (Ours) & 1552 & 111.34 & \textbf{36.15} & $\sim$ 57 \\
    \hline
  \end{tabular}
\end{table}

\begin{figure}[!t]
  \centering 
  \begin{subfigure}{0.49\textwidth}
      \centering
      \includegraphics[width=\textwidth]{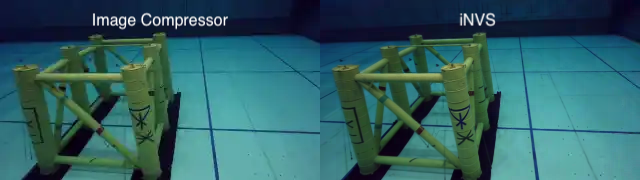}
     \caption{Reconstructed images for T1.}
     \label{fig:output_comparison}
  \end{subfigure}
  \begin{subfigure}{0.49\textwidth}
      \centering
      \includegraphics[width=\textwidth]{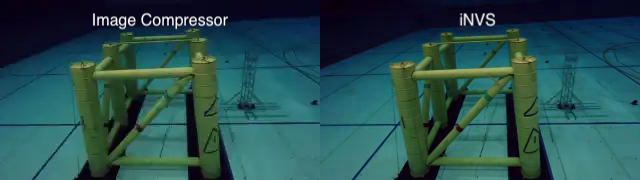}
     \caption{Reconstructed images for T2.}
     \label{fig:output_comparison_new_struct}
  \end{subfigure}\hfill
  \caption{Visual results showing reconstruction quality for T1 and T2. The image on the left is compressed/decompressed by JPEG-XL, the one on the right is reconstructed using NVSPrior+iNVS+JPEG-XL.}
\end{figure}

The results demonstrate that our NVSPrior+iNVS approach achieves the best overall performance among all methods. It achieves a significantly higher compression ratio than WebP and JPEG-XL while maintaining a higher PSNR. iNVS with WebP achieves the highest compression ratio, which is 2.90 times higher than WebP and 4.67 times higher than JPEG-XL. 

iNVS with JPEG-XL achieves the highest PSNR of 36.15~dB, which is 2.85~dB higher than WebP and 2.58~dB higher than JPEG-XL. In Fig.~\ref{fig:output_comparison}, we observe the reconstructed image from iNVS is clearer and sharper than the image compressed and decompressed by classical methods. The results demonstrate that our iNVS technique is more effective than classical compression methods for real-time image transmission over limited-bandwidth acoustic links.

Despite strong results on standard learned image compression benchmarks~\cite{compressai, mlic}, Mean \& Scale Hyperprior and MLIC++ underperform in our underwater setting. We attribute this to (i) limited domain-specific training data, which constrains learned priors and weakens context modeling, and (ii) a resolution mismatch—MLIC++ is primarily evaluated at higher resolutions, whereas our inputs are small, diminishing the benefit of its multi-reference context modules. In contrast, our NVS-based approach exploits a scene-specific 3D prior learned across viewpoints, assimilating 3D information more efficiently from modest training datasets and generalizing across repeat surveys; under these data-scarce, low-resolution, and underwater-degraded conditions, it achieves better rate–distortion performance.

\begin{figure*}[!t]
  \centering 
  \begin{subfigure}{0.49\textwidth}
      \includegraphics[width=\textwidth]{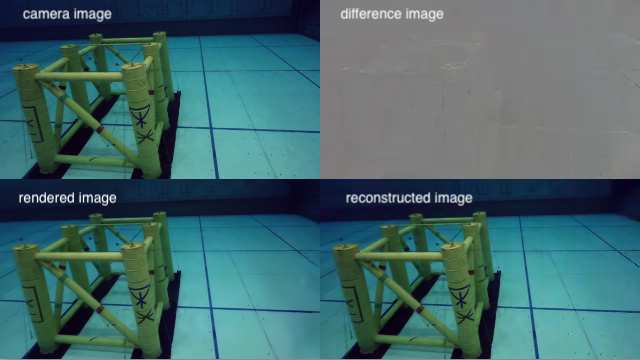}
     \caption{Result using NVSPrior+iNVS.}
     \label{fig:iNVS}
  \end{subfigure}\hfill
  \begin{subfigure}{0.49\textwidth}
      \includegraphics[width=\textwidth]{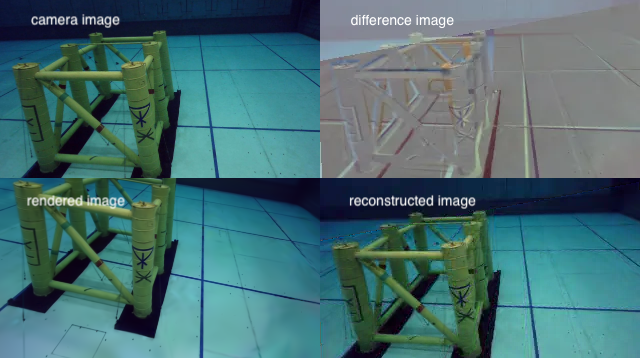}
     \caption{Result using NVSPrior+Affine.}
     \label{fig:posenet-3dgs}
  \end{subfigure}\hfill
  \begin{subfigure}{0.49\textwidth}
      \includegraphics[width=\textwidth]{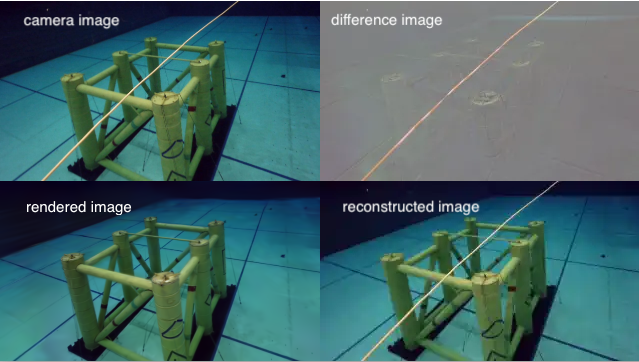}
     \caption{Result using NVSPrior+iNVS when an unmodeled safety line is present in the scene.}
     \label{fig:tether-occlusion}
  \end{subfigure}\hfill
  \begin{subfigure}{0.49\textwidth}
      \includegraphics[width=\textwidth]{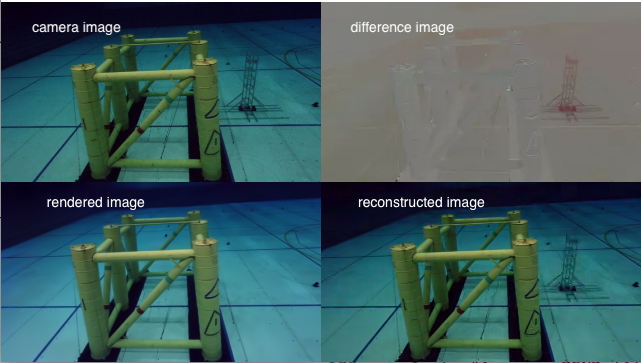}
     \caption{Result using NVSPrior+iNVS when a new structure is present in the scene.}
     \label{fig:new-struct-iNVS}
  \end{subfigure}\hfill
  \caption{Visualization of the compression performance using NVSPrior with either the iNVS (a, c, d) or the Affine approach (b). In each of the subfigures, we present the (i) camera image, (ii) rendered image at the estimated latent representation, (iii) the difference between the two images and (iv) the final reconstructed image. The visible artifacts in (b) arise from pose estimation errors and the limitations of affine transformation.}
\end{figure*}

We also find the overall performance of NVSPrior+Affine is worse than both NVSPrior+iNVS and classical codecs. This is likely because the affine transformation method assumes a 2D scene geometry and introduces visual artifacts. As illustrated in Fig.~\ref{fig:posenet-3dgs}, the latent representation estimated by PoseLSTM often results in significant misalignment between the rendered and camera images, leading to larger residuals. This misalignment cannot be fully corrected by affine warping, which introduces visual artifacts, thereby increasing the entropy of the difference image. This results in a much larger compressed size using Affine than using iNVS. This underscores the effectiveness of the latent representation optimization by iNVS.

Using our approach, the average transmitted data size is 1.2~kB, allowing approximately 10 frames per second to be sent over a 100~kbps acoustic link. Although our method involves an additional optimization step, iNVS remains computationally efficient, with a runtime of approximately 62~ms per frame. This is largely due to effective initialization from the previous frame’s optimized latent representation, which enables rapid convergence of the BFGS optimizer. These properties make NVSPrior+iNVS a practical and scalable solution for real-time image transmission in bandwidth-constrained underwater inspection scenarios.

\subsection{Robustness to Novel Objects in the Scene}

In inspection missions, it is common to encounter changes in the scene with time, such as the presence of additional structures or objects (e.g. fish, biological growth, corrosion, etc). We evaluate the robustness of our NVSPrior+iNVS technique to novel objects in the scene using dataset T2, in a similar manner as described above.

We test the performance of iNVS on two examples, representing novel objects commonly encountered in inspection missions. The first is a thin yellow safety line that moves with the vehicle and appears on the camera, as shown in Fig.~\ref{fig:tether-occlusion}. The second is a stationary metallic object with dimensions approximately 1.0~m~$\times$~0.25~m~$\times$~0.25~m, as shown in Fig.~\ref{fig:new-struct-iNVS}. We find that NVSPrior+iNVS handles both types of novel objects well. The average compressed data size with the presence of these objects is 1.65~kB, allowing us to transmit about 7 frames per second over a 100~kbps acoustic link. The results are summarized in Table~\ref{tab:t2_results}.

\begin{table}[!h]
  \renewcommand{\arraystretch}{1.3}
  \caption{Quantitative results on the T2 dataset, averaged over 1000 \(320\times180\) RGB images.}
  \label{tab:t2_results}
  \centering
  \begin{tabular}{l|c|c|c|c}
    \hline
    Method & Size (bytes) $\downarrow$ & Ratio $\uparrow$ & PSNR (dB) $\uparrow$ & Time (ms) $\downarrow$\\
    \hline\hline
    WebP & 3655 & 47.28 & 33.43 & $\sim$ 6 \\
    JPEG-XL & 5827  & 29.66 & 33.86 & $\sim$ \textbf{1} \\
    Mean \& Scale Hyperprior~\cite{mbt18} &  13794 & 12.53 & 35.35 & $\sim$22 \\
    MLIC++~\cite{mlic} & 4260 & 40.57 & 30.93 & $\sim$ 128\\
    NVSPrior+Affine+WebP~\cite{rajat_oceans_2024} & 4452 & 38.81 & 31.83 & $\sim$ 92 \\
    NVSPrior+Affine+JPEG-XL~\cite{rajat_oceans_2024} & 4629 & 37.33  & 31.37 & $\sim$ 86 \\
    NVSPrior+iNVS+WebP (Ours) & \textbf{1651} & \textbf{104.66} & 35.32 & $\sim$ 125 \\
    NVSPrior+iNVS+JPEG-XL (Ours) & 2073 & 83.36 & \textbf{35.55} & $\sim$ 119 \\
    \hline
  \end{tabular}
\end{table}

We find that even with the presence of novel objects in the scene, our NVS-prior approach remains the most bitrate-efficient and also attains the highest reconstruction quality. NVSPrior+iNVS+WebP achieves the smallest transmitted size, improving compression by 2.21 times over WebP and 3.53 times over JPEG-XL. NVSPrior+iNVS+JPEG-XL yields the highest PSNR, exceeding WebP by 2.12~dB and JPEG-XL by 1.69~dB, while still reducing size to 2073~bytes.

Learned baselines do not perform well in this setting: the Mean \& Scale Hyperprior achieves 35.35~dB PSNR but at a much higher bitrate, and MLIC++ produces lower quality with a larger size than our methods. Relative to the prior Affine variant, iNVS improves both rate and distortion, underscoring the importance of latent refinement rather than 2D warping.

Compared with T1, iNVS exhibits a modest degradation on T2 in compression efficiency, reconstruction quality, and runtime, which we attribute primarily to two factors.  First, the presence of a novel object in the scene increases the difficulty for the trained estimator to provide a good initialization. For the initial frames, we rely on the PoseLSTM estimator, as the rendered image from the previous frame is determined to be too different from the current frame, as shown by its MSE which is greater than the threshold. Hence, the $I_{\text{diff}}$ energies for the first few frames are larger due to less accurate pose initialization. Second, the presence of the novel object increases the entropy of $I_{\text{diff}}$, resulting in a larger compressed size. Even so, the NVS-prior approach remains decisively better than classical codecs and learned baselines on both rate and distortion in this data-scarce, low-resolution, underwater setting.

\subsection{Compression Performance on Real-world Datasets}

We evaluate the performance of our NVSPrior+iNVS method on both the IUI3-RedSea dataset and the Torpedo Boat Wreck dataset, using the same configuration and evaluation protocols as in our controlled experiments. Since WebP consistently outperforms JPEG-XL in our earlier tests, we use WebP both to compress our residual images and as the classical codec benchmark. The NVSPrior+Affine baseline is omitted, as it was consistently outperformed by our iNVS method.

\begin{figure}[t]
  \centering
  % Column 1: Sample 1
  \begin{subfigure}{0.32\columnwidth}
    \includegraphics[width=\linewidth]{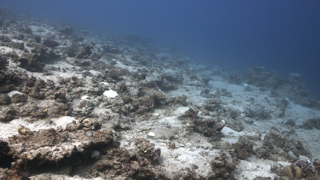}
    \par\vspace{0.3em}
    \includegraphics[width=\linewidth]{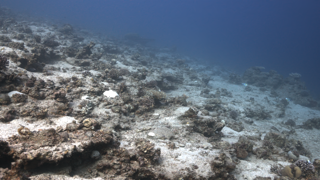}
    \caption*{Sample 1}
  \end{subfigure}%
  \hfill
  % Column 2: Sample 2
  \begin{subfigure}{0.32\columnwidth}
    \includegraphics[width=\linewidth]{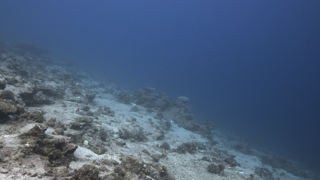}
    \par\vspace{0.3em}
    \includegraphics[width=\linewidth]{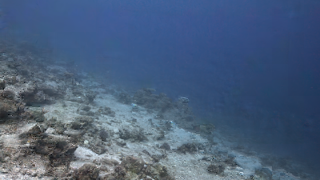}
    \caption*{Sample 2}
  \end{subfigure}%
  \hfill
  % Column 3: Sample 3
  \begin{subfigure}{0.32\columnwidth}
    \includegraphics[width=\linewidth]{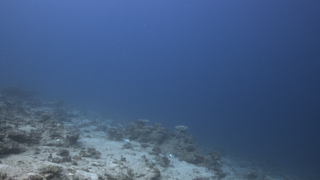}
    \par\vspace{0.3em}
    \includegraphics[width=\linewidth]{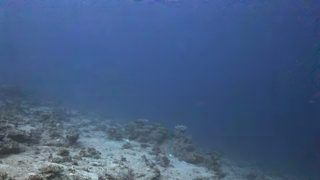}
    \caption*{Sample 3}
  \end{subfigure}

  \caption{Representative qualitative comparison for three samples from the IUI3-RedSea dataset. For each sample (column), the top image shows the raw camera frame and the bottom image shows the corresponding NVSPrior+iNVS reconstruction.}
  \label{fig:qualitative_coral}
\end{figure}

Table~\ref{tab:coral_results} summarizes the quantitative results for the IUI3-RedSea dataset. NVSPrior+iNVS achieves the best rate–distortion performance, reducing the transmitted size to 1670~bytes and attaining a PSNR of 33.46~dB. This represents a 2.49$\times$ higher compression ratio and a 0.97~dB improvement in PSNR over WebP. Learned baselines, including Mean \& Scale Hyperprior and MLIC++, require substantially more bytes while delivering lower reconstruction quality, likely due to limited, domain-specific training data available at this low resolution. Qualitatively, as shown in Fig.~\ref{fig:qualitative_coral}, our method preserves scene structure with minimal degradation.

\begin{table}[!h]
  \renewcommand{\arraystretch}{1.3}
  \caption{Quantitative results on the IUI3-RedSea dataset, averaged over 29 \(320\times180\) RGB images.}
  \label{tab:coral_results}
  \centering
  \begin{tabular}{l|c|c|c|c}
    \hline
    Method & Size (bytes) $\downarrow$ & Ratio $\uparrow$ & PSNR (dB) $\uparrow$ & Time (ms)\\
    \hline\hline
    WebP & 4151 & 41.63 & 32.49 & $\sim$ \textbf{6} \\
    Mean \& Scale Hyperprior~\cite{mbt18} & 13506 & 12.79 & 23.64 & $\sim$127 \\
    MLIC++~\cite{mlic} & 7346 & 23.52 & 25.07 & $\sim$131\\
    NVSPrior+iNVS (Ours) & \textbf{1670} & \textbf{103.47} & \textbf{33.46} & $\sim$ 312 \\
    \hline
  \end{tabular}
\end{table}

\begin{figure*}[t!]
  \centering
  % First sample image row
  \begin{subfigure}{\textwidth}
    \centering
    \includegraphics[width=0.19\textwidth]{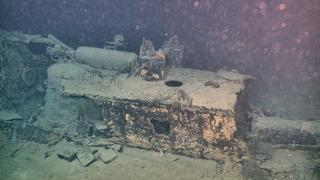}
    \includegraphics[width=0.19\textwidth]{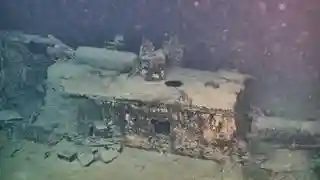}
    \includegraphics[width=0.19\textwidth]{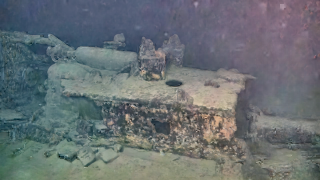}
    \includegraphics[width=0.19\textwidth]{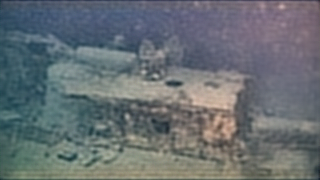}
    \includegraphics[width=0.19\textwidth]{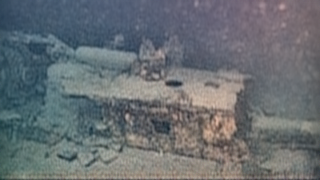}
  \end{subfigure}

  \vspace{0.5em} % Space between rows

  % Second sample image row
  \begin{subfigure}{\textwidth}
    \centering
    \includegraphics[width=0.19\textwidth]{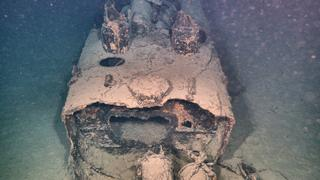}
    \includegraphics[width=0.19\textwidth]{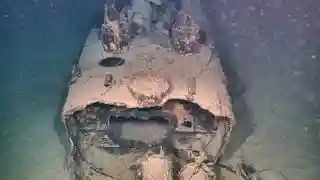}
    \includegraphics[width=0.19\textwidth]{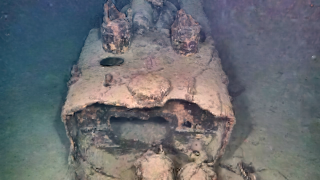}
    \includegraphics[width=0.19\textwidth]{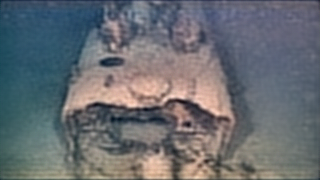}
    \includegraphics[width=0.19\textwidth]{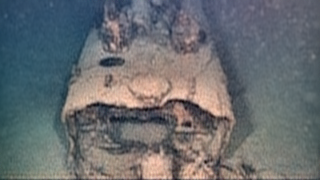}
  \end{subfigure}

  \vspace{0.5em}

  % Legend aligned with image columns
  {\footnotesize
    \begin{minipage}{\textwidth}
      \centering
      \begin{tabular}{@{\hspace{0.0\textwidth}}c@{\hspace{0.08\textwidth}}c@{\hspace{0.07\textwidth}}c@{\hspace{0.07\textwidth}}c@{\hspace{0.07\textwidth}}c@{}}
        \textbf{Ground Truth} & \textbf{Classic Codec} & \textbf{NVSPrior+iNVS} & \textbf{Mean \& Scale Hyperprior} & \textbf{MLIC++} \\
        (Camera) & (WebP) & (Ours) & \cite{mbt18} & \cite{mlic}
      \end{tabular}
    \end{minipage}
  }

  \caption{Qualitative comparison of compression results for two representative images from the Torpedo Boat Wreck dataset. For each sample, we show (from left to right): the ground truth camera image, the reconstruction using a classical codec (WebP), our NVSPrior+iNVS method, the Mean \& Scale Hyperprior~\cite{mbt18}, and MLIC++~\cite{mlic}. NVSPrior+iNVS provides more faithful preservation of scene structure and details, with fewer artifacts and higher visual fidelity under challenging underwater conditions.}
  \label{fig:torpedo_qualitative_comparison}
\end{figure*}

For the Torpedo Boat Wreck dataset also (see Table~\ref{tab:torpedo_results}), NVSPrior+iNVS demonstrates strong performance, requiring only 1996~bytes on average to transmit a frame, compared to 4422~bytes for WebP, 9789~bytes for Mean \& Scale Hyperprior, and 6232~bytes for MLIC++. The reliability of PSNR as a useful metric is compromised by the presence of marine snow, as this random noise reduces the informativeness of pixel-wise similarity metrics. Although our PSNR is indicated to be lower than that of WebP and MLIC++ when considering the real data as a baseline, visual inspection (Fig.~\ref{fig:torpedo_qualitative_comparison}) shows that our method maintains higher visual fidelity and fewer artifacts across challenging underwater scenes.

These results highlight the robustness and effectiveness of NVSPrior+iNVS in real-world, data-scarce,and noisy underwater imaging scenarios.

\begin{table}[!h]
  \renewcommand{\arraystretch}{1.3}
  \caption{Quantitative results on the Torpedo Boat Wreck dataset, averaged over 147 \(320\times180\) RGB images.}
  \label{tab:torpedo_results}
  \centering
  \begin{tabular}{l|c|c|c|c}
    \hline
    Method & Size (bytes) $\downarrow$ & Ratio $\uparrow$ & PSNR (dB) $\uparrow$ & Time (ms)\\
    \hline\hline
    WebP & 4422 & 39.08 & \textbf{30.75} & $\sim$ \textbf{6} \\
    Mean \& Scale Hyperprior~\cite{mbt18} & 9789 & 17.65 & 27.93 & $\sim$15 \\
    MLIC++~\cite{mlic} & 6232 & 27.73 & 29.34 & $\sim$125\\
    NVSPrior+iNVS (Ours) & \textbf{1996} & \textbf{86.57} & 28.96 & $\sim$ 252 \\
    \hline
  \end{tabular}
\end{table}

\section{Discussion}
\label{sec:discussion}
While our proposed NVSPrior+iNVS framework demonstrates strong compression efficiency and robustness in both controlled and real-world underwater datasets, several important limitations and practical considerations remain for real-world deployment.

\subsection{Runtime and Edge Deployment}
\begin{figure}
    \centering
    \includegraphics[width=0.48\linewidth]{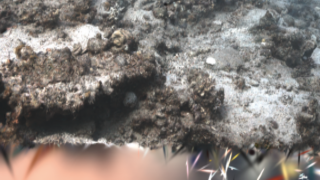}
    \includegraphics[width=0.48\linewidth]{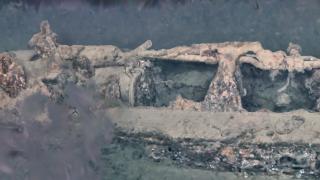}
    \caption{Example image rendered by the 3DGS model. Left: example images rendered with color artifacts at boundrys for model trained on the IUI3-RedSea dataset due to limited training views.  Right: example image rendered with floater with model trained on the torpedo boat dataset due to the lighting inconsistency and marine snow.}
    \label{fig:nvs-artifacts}
\end{figure}

A key limitation is computational performance. In real-world datasets, the per-frame runtime of iNVS is significantly higher—about four to six times longer than in controlled settings (e.g., T1). This increase is primarily due to less reliable pose initialization and the need for more optimization iterations, as reliable pose estimators or previous-frame reuse are often unavailable with public datasets. Moreover, limitations in real datasets such as limited viewpoints in the IUI3-RedSea dataset and marine snow and lighting inconsistency due to use of flash introduce artifacts in the 3DGS renderings (Fig.~\ref{fig:nvs-artifacts}), raising the residual energy in \(I_{\text{diff}}\) and making gradient-based refinement harder. Furthermore, when tested on the Jetson Orin NX (16~GB), an edge device, the runtime approximately doubles compared to our workstation results. This highlights the need for further code optimization and possibly model simplification to enable real-time deployment on resource-constrained platforms.

\subsection{Limitations of Evaluation Protocol}
Another limitation arises from the nature of the publicly available evaluation datasets. These datasets only provide images from a single mapping trial, and our experiments select test images from the same trial as used for training the NVS prior. Consequently, although our method shows robustness to certain in-the-wild challenges—such as changing illumination and water conditions (e.g., in the Torpedo Boat Wreck dataset, aided by image preprocessing)—the model’s generalization to evolving environments across repeated inspection runs is not fully verified. 

\subsection{Implications for Practical Deployment}
Our prior field trial experiences indicate that in real world, the frequency of revisiting and updating the scene prior must be tailored to the dynamics of the operational environment. For example, in warm, shallow waters such as those around Singapore, rapid marine growth can change scene appearance within weeks, necessitating more frequent mapping runs to maintain compression effectiveness. If the environment evolves faster than the revisit rate, the prior may become outdated, resulting in reduced compression ratios and reconstruction quality. Verifying this empirically will require the collection of new field datasets that capture environmental changes across multiple inspection runs at the same site—data that is not currently available publicly.

\subsection{Directions for Future Work}
To address these limitations, future work will focus on: (i) optimizing the iNVS pipeline for faster inference and deployment on embedded hardware; (ii) collecting and evaluating field datasets with repeated surveys over time to rigorously assess robustness to environmental evolution; and (iii) improving the robustness of iNVS in handling dynamic changes in underwater environments. These efforts will be essential for translating the demonstrated promise of NVSPrior+iNVS into reliable, real-time visual feedback for practical underwater inspection missions.

\section{Conclusion}
\label{sec:conclusion}
In this paper, we proposed NVSPrior+iNVS, a novel image compression approach for real-time image transmission over underwater acoustic links in inspection missions. Our method explicitly incorporated scene-specific knowledge through a pretrained NVS model, allowing it to reconstruct camera views from compact latent representations with high accuracy—even when training data was limited. Through extensive evaluations on both controlled and real-world underwater datasets, we demonstrated that this scene-aware design enabled NVSPrior+iNVS consistently outperforms classical codecs and end-to-end learned image compression models, achieving higher compression ratio and better reconstruction quality. We further analyzed design choices and found that MSE loss, BFGS optimization, and prior-frame initialization offered the best trade-off between compression performance and runtime. While the runtime increased under challenging real-world conditions due to degraded NVS priors, our method remained robust and practical for near-real-time transmission, even in the presence of novel objects or occlusions.

Looking forward, we aim to further enhance our method's adaptability to dynamic underwater environments, extend it for real-time video streaming, and design specialized compressors for sparse residual data. These advances will help enable practical, high-quality visual feedback in future underwater inspection missions and other bandwidth-constrained applications.

% \end{blueblock}

\section*{Acknowledgment}
This research project is supported by A*STAR under its RIE2020 Advanced Manufacturing and Engineering (AME) Industry Alignment Fund - Pre-Positioning (IAF-PP) Grant No. A20H8a0241.

\ifCLASSOPTIONcaptionsoff
  \newpage
\fi

\printbibliography
% \bibliographystyle{IEEEtran}
% \bibliography{references.bib}

\end{document}